\documentclass[a4paper, oneside, 12pt]{scrartcl}

\usepackage[utf8x]{inputenc}
\usepackage[T1]{fontenc}
\usepackage[english]{babel}

\makeatletter
\DeclareOldFontCommand{\rm}{\normalfont\rmfamily}{\mathrm}
\DeclareOldFontCommand{\sf}{\normalfont\sffamily}{\mathsf}
\DeclareOldFontCommand{\tt}{\normalfont\ttfamily}{\mathtt}
\DeclareOldFontCommand{\bf}{\normalfont\bfseries}{\mathbf}
\DeclareOldFontCommand{\it}{\normalfont\itshape}{\mathit}
\DeclareOldFontCommand{\sl}{\normalfont\slshape}{\@nomath\sl}
\DeclareOldFontCommand{\sc}{\normalfont\scshape}{\@nomath\sc}
\makeatother

\usepackage[a4paper,  centering, bindingoffset=0mm, inner=25mm, outer=25mm, top=26mm, bottom=36mm, heightrounded]{geometry}

\usepackage{graphicx}
\usepackage{float}
\usepackage{cite}

\usepackage{amsmath}	
\usepackage{amssymb}
\usepackage{amsfonts}
\usepackage{mathrsfs}
\usepackage{mathtools}
\usepackage{amsthm}
\usepackage{bm}
\usepackage{esint}     

\usepackage{braket}
\usepackage{slashed}
\newcommand{\D}{{\cal D}}

\usepackage[font=small,labelfont=bf, width=.95\textwidth]{caption} 

\usepackage{verbatim}
\usepackage{textcomp}
\usepackage{enumitem}

\usepackage[toc]{appendix}

\usepackage{layout}
\usepackage{microtype}
\usepackage{bookmark}
\usepackage{color}
\usepackage[]{xcolor}

\newcommand{\dd}{\mathrm{d}}
\newcommand{\hdd}{\hat{\mathrm{d}}}
\newcommand{\ii}{\mathrm{i}} 
\newcommand{\hdelta}{\hat{\delta}}

\makeatletter
\newcommand{\subalign}[1]{%
	\vcenter{%
		\Let@ \restore@math@cr \default@tag
		\baselineskip\fontdimen10 \scriptfont\tw@
		\advance\baselineskip\fontdimen12 \scriptfont\tw@
		\lineskip\thr@@\fontdimen8 \scriptfont\thr@@
		\lineskiplimit\lineskip
		\ialign{\hfil$\m@th\scriptstyle##$&$\m@th\scriptstyle{}##$\crcr
			#1\crcr
		}%
	}
}
\makeatother
\usepackage{tikz}
\usepackage{circuitikz}

\usepackage{tikz-feynman}
\tikzfeynmanset{warn luatex=false}\usepackage{tikz-feynman}
\tikzfeynmanset{warn luatex=false}

\usetikzlibrary{decorations.pathmorphing}
\usetikzlibrary{shapes.geometric}
\usetikzlibrary{matrix, calc, arrows}
\tikzset{snake it/.style={decorate, decoration=snake}}
\usetikzlibrary{decorations.markings}

\tikzset{%
	dots/.style args={#1per #2}{%
		line cap=round,
		dash pattern=on 0 off #2/#1
	}
}

\definecolor{unamblue}{rgb}{0.0, 0.0, 0.0}

\usepackage[]{hyperref}
\hypersetup{
	colorlinks=true,%
	citecolor=unamblue,%
	filecolor=unamblue,%
	linkcolor=unamblue,%
	urlcolor=unamblue,
	bookmarksnumbered=true,     
	bookmarksopen=true,         
	bookmarksopenlevel=1,       
	pdfstartview=Fit,           
	pdfpagemode=UseOutlines,
	pdfpagelayout=TwoPageRight
}

\usepackage{graphicx}
\usepackage{tikz}
\usepackage{slashed}
\usepackage{epstopdf}
\usepackage{verbatim}	
\usepackage{blkarray}
\usepackage{ytableau}

\newcommand{\ba }{\begin{equation}\begin{aligned}}
\newcommand{\ea }{\end{aligned}\end{equation}}
\newcommand{\deltahat}{\hat{\delta}}
\newcommand{\bu}{\bar{u}}

\usepackage[]{todonotes}
\usepackage{subcaption}
\usepackage{verbatim}
\usepackage{textcomp}
\usepackage{enumitem}

\usepackage[toc]{appendix}

\usepackage{layout}
\usepackage{microtype}
\usepackage{bookmark}
\usepackage{color}
\usepackage[]{xcolor}

\makeatother
\usepackage{tikz}
\usetikzlibrary{decorations.pathmorphing}
\usetikzlibrary{shapes.geometric}
\usetikzlibrary{matrix, calc, arrows}
\tikzset{snake it/.style={decorate, decoration=snake}}
\usetikzlibrary{decorations.markings}

\tikzset{%
	dots/.style args={#1per #2}{%
		line cap=round,
		dash pattern=on 0 off #2/#1
	}
}

\definecolor{unamblue}{rgb}{0.0, 0.0, 0.0}

\usepackage[]{hyperref}
\hypersetup{
	colorlinks=true,%
	citecolor=unamblue,%
	filecolor=unamblue,%
	linkcolor=unamblue,%
	urlcolor=unamblue,
	bookmarksnumbered=true,     
	bookmarksopen=true,         
	bookmarksopenlevel=1,       
	pdfstartview=Fit,           
	pdfpagemode=UseOutlines,
	pdfpagelayout=TwoPageRight
}

\usepackage{graphicx}
\usepackage{tikz}
\usepackage{slashed}
\usepackage{epstopdf}
\usepackage{verbatim}	
\usepackage{blkarray}
\usepackage{ytableau}
\usepackage{eufrak}

\usepackage[]{todonotes}
\usepackage{subcaption}

\newcommand{\s}{{\cal S}}

\newcommand{ \vkone} {\begin{tikzpicture}[thick]
	\path [dashed, draw]
	(-1,-1) -- (0,-1);
	\path [draw,very thick]
	(0,-1) -- (1,-1) node[right]
	{$q_\nu(\omega)$};
	\path [ draw,snake it]
	(0,-1) -- (0,-2)node[below]{$A^{a}_{\mu}(k)$};
	\draw[->,>=stealth] (-0.5,-1.2) -- (-0.5,-1.8) node[midway, left]
	{$k$};
	\coordinate (A) at (0,-1);
	\filldraw (A) circle (1.5pt);
	\end{tikzpicture}
}

\newcommand{\vkzeroGR} {
\begin{tikzpicture}[thick]
	\path [dashed, draw]
	(-1,-1) -- (0,-1);
	\path [dashed, draw]
	(0,-1) -- (1,-1) ;
	\path [ draw,double,snake it]
	(0,-1) -- (0,-2)node[below]{$h_{\mu\nu}(k)$};
	\draw[->,>=stealth] (-0.5,-1.2) -- (-0.5,-1.8) node[midway, left]
	{$k$};
	\coordinate (A) at (0,-1);
	\filldraw (A) circle (1.5pt);
	\end{tikzpicture}
}

\newcommand{ \vkoneGR} {\begin{tikzpicture}[thick]
	\path [dashed, draw]
	(-1,-1) -- (0,-1);
	\path [draw,very thick]
	(0,-1) -- (1,-1) node[right]
	{$q_\alpha(\omega)$};
	\path [ draw,double,snake it]
	(0,-1) -- (0,-2)node[below]{$h_{\mu\nu}(k)$};
	\draw[->,>=stealth] (-0.5,-1.2) -- (-0.5,-1.8) node[midway, left]
	{$k$};
	\coordinate (A) at (0,-1);
	\filldraw (A) circle (1.5pt);
	\end{tikzpicture}
}

\newcommand{ \vright} {\begin{tikzpicture}[thick]
	\path [dashed, draw]
	(-1,-1) -- (0,-1);
	\path [draw,very thick,red]
	(0,-1) -- (1,-1) node[right,black]
	{$\beta^\sigma(\omega)$};
	\path[](0.2,-1) -- (0.6,-1) node[color=red,currarrow,scale=1,xscale=1]{};
	\path [ draw,snake it]
	(0,-1) -- (0,-2)node[below]{$A^{a}_{\mu}(k)$};
	\draw[->,>=stealth] (-0.5,-1.2) -- (-0.5,-1.8) node[midway, left]
	{$k$};
	\coordinate (A) at (0,-1);
	\filldraw (A) circle (1.5pt);
	\end{tikzpicture}
}

\newcommand{ \vrightAdj} {\begin{tikzpicture}[thick]
	\path [dashed, draw]
	(-1,-1) -- (0,-1);
	\path [draw,very thick,red]
	(0,-1) -- (1,-1) node[right,black]
	{$\beta^{\mu}(\omega)$};
	\path[](0.2,-1) -- (0.6,-1) node[color=red,currarrow,xscale=1,sloped,scale=1,pos=.7]{};
	\path [ draw,snake it]
	(0,-1) -- (0,-2)node[below]{$\varphi^{a\alpha}(k)$};
	\draw[->,>=stealth] (-0.5,-1.2) -- (-0.5,-1.8) node[midway, left]
	{$k$};
	\coordinate (A) at (0,-1);
	\filldraw (A) circle (1.5pt);
	\end{tikzpicture}
}

\newcommand{ \vrightAdjd} {\begin{tikzpicture}[thick]
	\path [dashed, draw]
	(-1,-1) -- (0,-1);
	\path [draw,very thick,blue]
	(0,-1) -- (1,-1) node[right,black]
	{$\eta^{\mu}(\omega)$};
	\path[](0.2,-1) -- (0.6,-1) node[color=blue,currarrow,xscale=1,sloped,scale=1,pos=.7]{};
	\path [ draw,snake it]
	(0,-1) -- (0,-2)node[below]{$\varphi^{a\alpha}(k)$};
	\draw[->,>=stealth] (-0.5,-1.2) -- (-0.5,-1.8) node[midway, left]
	{$k$};
	\coordinate (A) at (0,-1);
	\filldraw (A) circle (1.5pt);
	\end{tikzpicture}
}

\newcommand{\lotop}[3]
{\begin{tikzpicture}[thick]
	\draw[draw, snake it] (-0.5,0)--(-0.5,-1)node[below]
	{#1};
	\draw[<-,,<-=stealth, draw] (-0.8,-0.7)--(-0.8,-0.2) node[below left]
	{$k_1$};
	\draw[draw, snake it] (0.5,0)--(0.5,-1)node[below]
	{#2};
	\draw[<-,,<-=stealth, draw] (0.8,-0.7)--(0.8,-0.2) node[below right]
	{$k_2$};
	\path [dashed, draw] (0.5,0) --(1,0);
	\path [draw, ,very thick,#3]    (-0.5,0) --(.5,0);
	\path [dashed, draw]    (-1,0) --(-0.5,0);
	\coordinate (A) at (-0.5,0);   \filldraw (A) circle (2.pt);
	\coordinate (B) at (0.5,0);   \filldraw (B) circle (2.pt);
	\end{tikzpicture}
}

\newcommand{\lotopcol}[3]
{\begin{tikzpicture}[thick]
	\draw[draw, snake it] (-0.5,0)--(-0.5,-1)node[below]
	{#1};
	\draw[<-,,<-=stealth, draw] (-0.8,-0.7)--(-0.8,-0.2) node[below left]
	{$k_1$};
	\draw[draw, snake it] (0.5,0)--(0.5,-1)node[below]
	{#2};
	\draw[<-,,<-=stealth, draw] (0.8,-0.7)--(0.8,-0.2) node[below right]
	{$k_2$};
	\path [dashed, draw] (0.5,0) --(1,0);
	\path [draw, ,very thick,#3]    (-0.5,0) --(.5,0);
	\path[](-.2,0)--(.1,0) node[color=#3,currarrow,pos=.6, 
	xscale=1,
	sloped,
	scale=1]{};
	\path [dashed, draw]    (-1,0) --(-0.5,0);
	\coordinate (A) at (-0.5,0);   \filldraw (A) circle (2.pt);
	\coordinate (B) at (0.5,0);   \filldraw (B) circle (2.pt);
	\end{tikzpicture}
}

\newcommand{\lotopcolCross}[3]
{\begin{tikzpicture}[thick]
	\draw[draw, snake it] (-0.5,0)--(0.5,-1)node[below]
	{#1};
	\draw[<-,,<-=stealth, draw] (-0.8,-0.7)--(-0.8,-0.2) node[below left]
	{$k_2$};
	\draw[draw, snake it] (0.5,0)--(-0.5,-1)node[below]
	{#2};
	\draw[<-,,<-=stealth, draw] (0.8,-0.7)--(0.8,-0.2) node[below right]
	{$k_1$};
	\path [dashed, draw] (0.5,0) --(1,0);
	\path [draw, ,very thick,#3]    (-0.5,0) --(.5,0);
	\path[](-.2,0)--(.1,0) node[color=#3,currarrow,pos=.6, 
	xscale=1,
	sloped,
	scale=1]{};
	\path [dashed, draw]    (-1,0) --(-0.5,0);
	\coordinate (A) at (-0.5,0);   \filldraw (A) circle (2.pt);
	\coordinate (B) at (0.5,0);   \filldraw (B) circle (2.pt);
	\end{tikzpicture}
}

\newcommand{\lotopGR}[3]
{\begin{tikzpicture}[thick]
	\draw[draw, snake it,double] (-0.5,0)--(-0.5,-1)node[below]
	{#1};
	\draw[<-,,<-=stealth, draw] (-0.8,-0.7)--(-0.8,-0.2) node[below left]
	{$k_1$};
	\draw[draw, snake it,double] (0.5,0)--(0.5,-1)node[below]
	{#2};
	\draw[<-,,<-=stealth, draw] (0.8,-0.7)--(0.8,-0.2) node[below right]
	{$k_2$};
	\path [dashed, draw] (0.5,0) --(1,0);
	\path [draw, very thick,#3]    (-0.5,0) --(.5,0);
	\path [dashed, draw]    (-1,0) --(-0.5,0);
	\coordinate (A) at (-0.5,0);   \filldraw (A) circle (2.pt);
	\coordinate (B) at (0.5,0);   \filldraw (B) circle (2.pt);
	\end{tikzpicture}
}

\newcommand{\lotopGRspin}[3]
{\begin{tikzpicture}[thick]
	\draw[draw, snake it,double] (-0.5,0)--(-0.5,-1)node[below]
	{#1};
	\draw[<-,,<-=stealth, draw] (-0.8,-0.7)--(-0.8,-0.2) node[below left]
	{$k_1$};
	\draw[draw, snake it,double] (0.5,0)--(0.5,-1)node[below]
	{#2};
	\draw[<-,,<-=stealth, draw] (0.8,-0.7)--(0.8,-0.2) node[below right]
	{$k_2$};
	\path [dashed, draw] (0.5,0) --(1,0);
	\path [very thick,draw, #3]    (-0.5,0) --(0.5,0);
	\path[](-0.5,0)--(.1,0) node[color=#3,currarrow,pos=.8, 
	xscale=1,
	sloped,
	scale=1]{};
	\path [dashed, draw]    (-1,0) --(-0.5,0);
	\coordinate (A) at (-0.5,0);   \filldraw (A) circle (2.pt);
	\coordinate (B) at (0.5,0);   \filldraw (B) circle (2.pt);
	\end{tikzpicture}
}

\newcommand{\vGRspin}
{\begin{tikzpicture}[thick]
	\draw[draw, snake it,double] (-0.5,0)--(-0.5,-1)node[below]
	{$h_{\mu\nu}(k)$};
	\draw[<-,,<-=stealth, draw] (-0.8,-0.7)--(-0.8,-0.2) node[below left]
	{$k$};
	\path [very thick,draw, black]    (-0.5,0) --(0.5,0)node[right]{$\Psi^{\rho}(\omega) $};
	\path[] (-0.5,0) --(0.2,0) node[currarrow, xscale=1,sloped,scale=1,pos=.8]{};
	\path [dashed, draw]    (-1.5,0) --(-0.5,0);
	\coordinate (A) at (-0.5,0);   \filldraw (A) circle (2.pt);
	\end{tikzpicture}
}

\newcommand{\hhvertex }
{\begin{tikzpicture}[thick]
\path [dashed, draw]
	(-1,-1) -- (0,-1);
	\path [draw,dashed]
	(0,-1) -- (1,-1) ;
	\coordinate (A) at (0,-1);
	\filldraw (A) circle (1.5pt);
	\path [ draw,snake it,double ]
	(0,-1) -- (-1,-2)node[ left]{$h_{\mu\nu}$};
	\path [ draw,-> ]
	(0,-1.3) -- (-.5,-2)node[below left]{$k_1 $};

	\path [ draw,snake it,double ]
	(0,-1) -- (1,-2)node[right]{$h_{\alpha\beta}$};
	\path [ draw,-> ]
	(0,-1.3) -- (.5,-2)node[below right]{$k_2 $};
\end{tikzpicture}
}

\newcommand{\hhh}
{\begin{tikzpicture}[thick]
\path [dashed, draw]
	(-1,-1) -- (0,-1);
	\path [draw,dashed]
	(0,-1) -- (1,-1) ;
	\path [ draw,snake it,double ]
	(0,-1) -- (0,-2);
	\coordinate (A) at (0,-1);
	\coordinate (B) at (0,-2);

	\path [ draw,snake it,double ]
	(0,-2) -- (-1,-2.5)node[ left]{$h_{\mu\nu}$};
	\path [ draw,-> ]
	(0,-2.3) -- (-.5,-2.6)node[below left]{$k_1 $};
	
	\path [ draw,snake it,double ]
	(0,-2) -- (1,-2.5)node[right]{$h_{\alpha\beta}$};
	\path [ draw,-> ]
	(0,-2.3) -- (.5,-2.6)node[below right]{$k_2 $};
		\filldraw (A) circle (1.5pt);
	\filldraw (B) circle (2pt);
\end{tikzpicture}
}

\newcommand{\hhhF}
{\begin{tikzpicture}[thick]
		\path [draw]
		(-1,-1) -- (1,-1);
		\path [ draw,snake it,double ]
		(0,-1) -- (0,-2);
		\coordinate (A) at (0,-1);
		\coordinate (B) at (0,-2);

		\path [ draw,snake it,double ]
		(0,-2) -- (-1,-2.5)node[ left]{$h_{\mu\nu}$};
		\path [ draw,-> ]
		(0,-2.3) -- (-.5,-2.6)node[below left]{$k_1 $};
		\path [ draw,snake it,double ]
		(0,-2) -- (1,-2.5)node[right]{$h_{\alpha\beta}$};
		\path [ draw,-> ]
		(0,-2.3) -- (.5,-2.6)node[below right]{$k_2 $};
				\filldraw (A) circle (1.5pt);
		\filldraw (B) circle (2pt);
	\end{tikzpicture}
}
\newcommand{\hhhFclosed}
{\begin{tikzpicture}[thick]
		\draw (0,-0.5) circle (.5cm);
		\path [ draw,snake it,double ]
		(0,-1) -- (0,-2);
		\coordinate (A) at (0,-1);
		\coordinate (B) at (0,-2);
		\path [ draw,snake it,double ]
		(0,-2) -- (-1,-2.5)node[ left]{$h_{\mu\nu}$};
		\path [ draw,-> ]
		(0,-2.3) -- (-.5,-2.6)node[below left]{$k_1 $};
		\path [ draw,snake it,double ]
		(0,-2) -- (1,-2.5)node[right]{$h_{\alpha\beta}$};
		\path [ draw,-> ]
		(0,-2.3) -- (.5,-2.6)node[below right]{$k_2 $};
				\filldraw (A) circle (1.5pt);
		\filldraw (B) circle (2pt);
	\end{tikzpicture}
}

\newcommand{\nlotop}[5]{ \begin{tikzpicture}[thick]
	\draw[draw, snake it] (-2,0)--(-2,-1)node[below ]
	{#1};
	\draw[<-,,<-=stealth, draw] (-2.2,-0.7)--(-2.2,-0.2) node[ below left]
	{$k_1$};
	\draw[draw, snake it] (-1,0)--(-1,-1)node[below]
	{#2};
	\draw[<-,,<-=stealth, draw] (-1.2,-0.7)--(-1.2,-0.2) node[ below left]
	{$k_2$};
	\draw[draw, snake it] (0,0)--(0,-1)node[below]
	{#3};
	\draw[<-,,<-=stealth, draw] (-0.2,-0.7)--(-0.2,-0.2) node[ below left]
	{$k_3$};
	\path [dashed, draw]
	(-2.5,0) --(-1,0);
	\path [dashed, draw]
	(-1,0) --(0,0);
	\path [dashed, draw]
	(0,0) --(0.5,0);
	\path [draw, #4]
	(-2,0) --(-1,0);
	\coordinate (A) at (-1,0);   \filldraw (A) circle (2.pt);
	\coordinate (B) at (-2,0);   \filldraw (B) circle (2.pt);
	\coordinate (C) at (0,0);   \filldraw (C) circle (2.pt);
	\draw[#5]	(B) to [out=20,in=90] (C);
	\end{tikzpicture}}

\newcommand{\nlotopC}
{\begin{tikzpicture}[thick]
	\draw[draw, snake it] (-2,0)--(-2,-1);
	\draw[<-,,<-=stealth, draw] (-2.2,-0.7)--(-2.2,-0.2) node[below left]
	{$k_1$};
	\draw[draw, snake it] (-1,0)--(-1,-1);
	\draw[<-,,<-=stealth, draw] (-1.2,-0.7)--(-1.2,-0.2) node[below  left]
	{$k_2$};
	\draw[draw, snake it] (0,0)--(0,-1);
	\draw[<-,,<-=stealth, draw] (-0.2,-0.7)--(-0.2,-0.2) node[ below left]
	{$k_3$};
	\path [dashed, draw]
	(-2.5,0) --(-2,0);
	\path [thick, draw]
	(-2,0) --(-1,0);
	\path [dashed, draw]
	(-1,0) --(0,0);
	\path [dashed, draw]
	(0,0) --(0.5,0);

	\coordinate (A) at (-1,0);   \filldraw (A) circle (2.pt);
	\coordinate (B) at (-2,0);   \filldraw (B) circle (2.pt);
	\coordinate (C) at (0,0);   \filldraw (C) circle (2.pt);
	\coordinate (D) at (-.7,.4);  
	\coordinate (E) at (-1.3,.25);   
	
	\draw[very thick,red]	(B) to [out=20,in=90] (C);
	\draw[red](E) to (D) node[color=red,currarrow, 
	xscale=1,
	sloped,
	scale=1]{};
	\end{tikzpicture}
	}
		
\newcommand{\nlotopD}{ 
\begin{tikzpicture}[thick]
	\draw[draw, snake it] (-2,0)--(-2,-1);
	\draw[<-,,<-=stealth, draw] (-2.2,-0.7)--(-2.2,-0.2) node[below left]
	{$k_1$};
	\draw[draw, snake it] (-1,0)--(-1,-1);
	\draw[<-,,<-=stealth, draw] (-1.2,-0.7)--(-1.2,-0.2) node[below  left]
	{$k_2$};
	\draw[draw, snake it] (0,0)--(0,-1);
	\draw[<-,,<-=stealth, draw] (-0.2,-0.7)--(-0.2,-0.2) node[ below left]
	{$k_3$};
	\path [dashed, draw]
	(-2.5,0) --(-2,0);
	\path [thick, draw]
	(-2,0) --(-1,0);
	\path [dashed, draw]
	(-1,0) --(0,0);
	\path [dashed, draw]
	(0,0) --(0.5,0);

	\coordinate (D) at (-.7,.4);  
	\coordinate (E) at (-1.3,.25);   
	
	\draw[very thick,red]	(B) to [out=20,in=90] (C);
	
	\draw[red](E) to (D.south west) node[left,color=red,currarrow, 
	xscale=-1,
	sloped,
	scale=1]{};
	
		\coordinate (A) at (-1,0);   \filldraw (A) circle (2.pt);
	\coordinate (B) at (-2,0);   \filldraw (B) circle (2.pt);
	\coordinate (C) at (0,0);   \filldraw (C) circle (2.pt);

	\end{tikzpicture}
	}
		
\newcommand{\lotopfivePHone}[5]
{\begin{tikzpicture}[thick]
	\draw[draw, snake it] (-0.5,0)--(-0.5,-1)node[below]
	{#1};
	\draw[<-,,<-=stealth, draw] (-0.8,-0.7)--(-0.8,-0.2) node[below left]
	{$k_1$};
	\draw[draw, snake it] (0.5,0)--(0.5,-1)node[below]
	{#2};
	\draw[<-,,<-=stealth, draw] (0.8,-0.7)--(0.8,-0.2) node[below right]
	{$k_2$};
	\draw[draw, snake it] (1.5,0)--(1.5,-1)node[below]
	{#3};
	\draw[<-,,<-=stealth, draw] (1.8,-0.7)--(1.8,-0.2) node[below right]
	{$k_3$};
	\draw[draw, snake it] (2.5,0)--(2.5,-1)node[below]
	{#4};
	\draw[<-,,<-=stealth, draw] (2.8,-0.7)--(2.8,-0.2) node[below right]
	{$k_4$};
	\draw[draw, snake it] (3.5,0)--(3.5,-1)node[below]
	{#5};
	\draw[<-,,<-=stealth, draw] (3.8,-0.7)--(3.8,-0.2) node[below right]
	{$k_5$};

\coordinate (A) at (-0.5,0);   \filldraw (A) circle (2.pt);
	\coordinate (B) at (0.5,0);   \filldraw (B) circle (2.pt);
	\coordinate (C) at (1.5,0);   \filldraw (C) circle (2.pt);
	\coordinate (D) at (2.5,0);   \filldraw (D) circle (2.pt);
	\coordinate (E) at (3.5,0);   \filldraw (E) circle (2.pt);
	\draw[very thick]	(A) to (B);
	\draw[very thick]	(A) to [out=20,in=90] (C);
	\draw[very thick]	(A) to [out=20,in=90] (D);
	\draw[very thick]	(A) to [out=20,in=90] (E);
	\draw[dashed] (B) to (E);
	\draw[dashed](-1,0)to(-.5,0);
	\draw[dashed](3.5,0)to(4,0);
		
	\end{tikzpicture}
}

\newcommand{\lotopfivePHtwo}[5]
{\begin{tikzpicture}[thick]
	\draw[draw, snake it] (-0.5,0)--(-0.5,-1)node[below]
	{#1};
	\draw[<-,,<-=stealth, draw] (-0.8,-0.7)--(-0.8,-0.2) node[below left]
	{$k_1$};
	\draw[draw, snake it] (0.5,0)--(0.5,-1)node[below]
	{#2};
	\draw[<-,,<-=stealth, draw] (0.8,-0.7)--(0.8,-0.2) node[below right]
	{$k_2$};
	\draw[draw, snake it] (1.5,0)--(1.5,-1)node[below]
	{#3};
	\draw[<-,,<-=stealth, draw] (1.8,-0.7)--(1.8,-0.2) node[below right]
	{$k_3$};
	\draw[draw, snake it] (2.5,0)--(2.5,-1)node[below]
	{#4};
	\draw[<-,,<-=stealth, draw] (2.8,-0.7)--(2.8,-0.2) node[below right]
	{$k_4$};
	\draw[draw, snake it] (3.5,0)--(3.5,-1)node[below]
	{#5};
	\draw[<-,,<-=stealth, draw] (3.8,-0.7)--(3.8,-0.2) node[below right]
	{$k_5$};
	
	\path [draw,dashed] (-1,0) --(-.5,0);
	\path [draw,very thick] (0.5,0) --(1.5,0);
	\path [draw, ,very thick]    (-0.5,0) --(.5,0);
	\path [draw,very thick]    (1.5,0) --(2.5,0);
	\path [draw,dashed] (2.5,0) --(3.5,0);
	\path [draw,dashed] (3.5,0) --(4,0);

\coordinate (A) at (-0.5,0);   \filldraw (A) circle (2.pt);
	\coordinate (B) at (0.5,0);   \filldraw (B) circle (2.pt);
	\coordinate (C) at (1.5,0);   \filldraw (C) circle (2.pt);
	\coordinate (D) at (2.5,0);   \filldraw (D) circle (2.pt);
	\coordinate (E) at (3.5,0);   \filldraw (E) circle (2.pt);
	\draw[very thick]	(B) to [out=20,in=90] (E);
	
	\end{tikzpicture}
}

\newcommand{\lotopfivePHthree}[5]
{\begin{tikzpicture}[thick]
	\draw[draw, snake it] (-0.5,0)--(-0.5,-1)node[below]
	{#1};
	\draw[<-,,<-=stealth, draw] (-0.8,-0.7)--(-0.8,-0.2) node[below left]
	{$k_1$};
	\draw[draw, snake it] (0.5,0)--(0.5,-1)node[below]
	{#2};
	\draw[<-,,<-=stealth, draw] (0.8,-0.7)--(0.8,-0.2) node[below right]
	{$k_2$};
	\draw[draw, snake it] (1.5,0)--(1.5,-1)node[below]
	{#3};
	\draw[<-,,<-=stealth, draw] (1.8,-0.7)--(1.8,-0.2) node[below right]
	{$k_3$};
	\draw[draw, snake it] (2.5,0)--(2.5,-1)node[below]
	{#4};
	\draw[<-,,<-=stealth, draw] (2.8,-0.7)--(2.8,-0.2) node[below right]
	{$k_4$};
	\draw[draw, snake it] (3.5,0)--(3.5,-1)node[below]
	{#5};
	\draw[<-,,<-=stealth, draw] (3.8,-0.7)--(3.8,-0.2) node[below right]
	{$k_5$};
	
	\path [draw,dashed] (-1,0) --(-.5,0);
	
	\path [draw,dashed] (3.5,0) --(4,0);

\coordinate (A) at (-0.5,0);   \filldraw (A) circle (2.pt);
	\coordinate (B) at (0.5,0);   \filldraw (B) circle (2.pt);
	\coordinate (C) at (1.5,0);   \filldraw (C) circle (2.pt);
	\coordinate (D) at (2.5,0);   \filldraw (D) circle (2.pt);
	\coordinate (E) at (3.5,0);   \filldraw (E) circle (2.pt);
	\draw[very thick]	(A) to  (B);
	\draw[very thick]	(B) to  (C);
	\draw[very thick]	(C) to (D);
	\draw[very thick]	(D) to  (E);
	
	\end{tikzpicture}
}

\newcommand{\vnQED}{
	\begin{tikzpicture}[thick]
	\path [dashed, draw]
	(-1,-1) -- (0,-1);
	\draw[very thick] (0,-1) arc (-180:-280:1)node[right]
	{$q_{\alpha_n}(\omega_{n})$};
	\draw[very thick] (0,-1) arc (-210:-280:1)node[right]
	{$q_{\alpha_2}(\omega_{2})$};
	\path [dashed, very thick, dots=3 per .4cm, draw](.8,-.4) -- (.8,-0.1);
	\path [very thick,draw]
	(0,-1) -- (1,-1)node[right]
	{$q_{\alpha_1}(\omega_{1})$};
	\path [draw,snake it]
	(0,-1) -- (0,-2)node[below]{$A_{\mu}(k)$};
	\draw[->,>=stealth] (-0.5,-1.2) -- (-0.5,-1.8) node[midway, left]
	{$k$};
	\coordinate (A) at (0,-1);
	\filldraw (A) circle (1.5pt);
	\end{tikzpicture}
}

\newcommand{\ct}{\tilde c}

\title{\Huge Classical off-shell currents	
	\\}

\author{ \normalfont\normalsize Francesco Comberiati$^{a}$, Leonardo de la Cruz$^{a,b}$\\[2mm]
	\emph{\normalfont\small \em ${}^a$Dipartimento di Fisica e Astronomia ``Augusto Righi'', Universit\`a di Bologna}\\
	\emph{\normalfont\small \em and INFN Sezione di Bologna, via Irnerio 46, I-40126 Bologna, Italy}
	\\
	\emph{\normalfont\normalsize \em ${}^b$Institut de Physique Th\'eorique,  CEA, CNRS, Universit\'e Paris–Saclay,}\\
	\emph{\normalfont\normalsize \em F-91191 Gif-sur-Yvette cedex, France 	}
}
\date{%
			$\,$  
	\\[2\baselineskip]
	\normalfont\normalsize%
	\parbox{0.8\linewidth}{%
		{\bf \sf Abstract}.  We consider tree-level off-shell currents of two massive particles and $n$ massless bosons in the classical limit, which can be fused into the classical limit of $n+2$ scattering amplitudes. We show that dressing up the current with coherent wave-functions associated with the massive particles leads to the recently proposed  Worldline Quantum Field Theory (WQFT)  path integral. The currents thus constructed encode solutions of classical equations of motion so they can be applied to  contexts where the classical limit is relevant, including hard thermal loops. We give several examples of these currents in scalar, gauge and gravitational theories.            
	}
}

\begin{document}
\maketitle
\thispagestyle{empty}
\newpage
\tableofcontents

\section{Introduction} 
Off-shell currents are ordinarily used in efficient recursive evaluation of scattering amplitudes. These are computed from 
 off-shell currents by setting 
the external legs on-shell. Off-shell currents were famously employed by  Berends and Giele   \cite{Berends:1987me} to recursively calculate tree-level gluon scattering and they are routinely used in  matrix element generators \cite{Gleisberg:2008fv, Cafarella:2007pc, Lifson:2022mxf}.  Despite their success in practical calculations many of the appealing structures  of on-shell scattering amplitudes are  either not manifest 
or difficult to expose \cite{Britto:2012qi, Mastrolia:2015maa, Jurado:2017xut} in off-shell currents. Similarly, in the classical limit of scattering amplitudes\footnote{The classical limit of amplitudes has been received considerable attention in the last few years. The subject has been reviewed extensively in Refs.\cite{Kosower:2022yvp,Buonanno:2022pgc}.}  some of the on-shell features present before taking classical limit are not inherited.  Off-shell currents in the classical limit have appeared recently in Refs.\cite{Goldberger:2016iau,Goldberger:2017frp,Goldberger:2017ogt,Shen:2018ebu,Ben-Shahar:2021zww,Cristofoli:2020hnk}. 

In this paper we are concerned with off-shell 
currents in the classical limit from the perspective of  Kosower-Maybee-O'Connell (KMOC) \cite{Kosower:2018adc, Maybee:2019jus, delaCruz:2020bbn, Cristofoli:2021vyo, Aoude:2021oqj}, 
the Worldline Quantum Field Theory
\cite{Mogull:2020sak,Jakobsen:2021smu,Jakobsen:2021lvp, Jakobsen:2021zvh,Shi:2021qsb, Bastianelli:2021nbs,Wang:2022ntx, Jakobsen:2022psy} and their connection. 
Throughout,  we will focus on the study of the classical limit of the tree-level off-shell current
\begin{align}
\mathcal{A}_n
(p_1, p_2, k_1, \dots, k_n)\Big|_{k_i^2\ne 0,\, p_i^2=m^2} \, ,
\end{align} 
which, assuming that  $p_1$, $p_2$ are the momenta of two massive particles and $k_1, \dots, k_n$ are the momenta of massless gauge bosons, can be fused into a $n+2$ scattering amplitude. Unlike typical currents in Berends-Giele recursions here more than one leg is off-shell.  Similar objects   have appeared previously in QCD, see e.g. Ref.\cite{Schwinn:2005pi}. The  
$n=2$ current with spinning massive particles  is relevant in the context of ongoing discussions of higher spin gravitational Compton amplitudes in the classical limit
\cite{Chiodaroli:2021eug,Johansson:2019dnu, Bautista:2021wfy, Saketh:2022wap, Bern:2022kto, Kosmopoulos:2021zoq,Jakobsen:2022zsx,Cangemi:2022abk,Bautista:2022wjf,Bern:2020buy, Cristofoli:2021jas,Aoude:2022thd}.

The remaining of the paper is organized as follows. In Section \ref{off-shell-classical}  we  start with a general definition of a current and  show how its classical limit is obtained in WQFT. In Section \ref{computational-part} we calculate currents in QED and gravity.
In Section \ref{currents-and-HTLs} we apply our methods to compute Hard Thermal Loops, which can be understood as classical objects in the high temperature regime. Our conclusions are presented in Section \ref{conclusions}.

\section{Off-shell currents in the classical limit}
\label{off-shell-classical}
Let us consider
a theory which models massive scalar particles coupled with massless gauge bosons. Let $p=(p_1, p_2)$ and $k=(k_1, \dots, k_n)$ denote  tuples of outgoing momenta  of  massive and massless external legs, respectively.  We define the  $n+2$ off-shell current by
\begin{align}
\mathcal{A}_n^{I_1, \dots, I_n}
(p, k):= \hdelta^4(p_1+p_2+\sum_{i=1}^n k_i)
A_n^{I_1, \dots, I_n}
(p,k),
\label{general-def}
\end{align}
where,  without loss of generality 
the  massive scalars obey $p_i^2=m^2$, while for the massless particles $k_i^2\ne 0$. We have introduced the notation $\hdelta^n (x):=(2\pi)^n \delta (x)$ and for later use we define $\hdd^n x:= (2\pi)^n \dd^n x$. 
 The upper indices denote collectively color and/or Lorentz indices associated with the massless particles. 
\begin{figure}
	\centering
	\begin{tikzpicture}[thick, transform shape]
	\node[circle, fill=lightgray, draw] (c) at (0,0){$\mathcal A$};
	\draw[draw, snake it] (c.north west)--(-1.,1)node[left]
	{$\qquad k_1$};
	\draw[draw, snake it] (c.north east)--(1.,1)node[right]
	{$k_n$};
	\draw[draw](1.5,-0.30)-- (c.south east)node[right,    currarrow,
	pos=0.5, 
	xscale=1,
	sloped,
	scale=1]
	{};
	\draw[draw] (-1.5,-0.30)--(c.south west)node[left,    currarrow,
	pos=0.5, 
	xscale=-1,
	scale=1]{};
	\node(A) at (0,1){$\dots$};
		\node(B) at (-1.5,0){$p_1$};
	 \node(C) at (1.5,0){$p_2$};
	\end{tikzpicture}
	\caption{Off-shell current.  The blob represents a sum over tree-level Feynman diagrams.  Wavy lines represent off-shell outgoing massless gauge bosons.}
	\label{current-off-shell}
\end{figure}
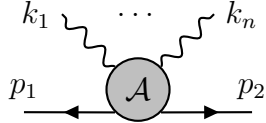
For our purposes the evaluation
of this current will be through Feynman diagrams (see Fig.\ref{current-off-shell}).  The 
current can be fused into an on-shell amplitude by
dressing  it with appropriate polarization vectors, imposing physical constraints such as transverse polarization, and setting all external legs on-shell,
namely
\begin{align}
\ii \mathcal T  \sim \mathcal \xi_{I_1} \cdots \xi_{I_n}\mathcal{A}_n^{I_1, \dots, I_n}
(p_1, p_2, k_1, \dots, k_n)\Big|_{k_i^2=0,p_i^2=m^2} \,.
\end{align}
\subsection{Classical limit \`a  la KMOC}
Suppose  that we are interested in the classical limit of Eq.\eqref{general-def}  understood as the limit $\hbar \to 0$ or more precisely as a Laurent expansion in powers of some dimensionless parameter $\xi$.  
 In the spirit of KMOC one
   would restore $\hbar$ into the current and perform dimensional analysis. One should also distinguish  the momenta of  massive scalar particles and massless gauge bosons. The latter being described by wavenumbers through the rescaling $k \to \hbar k$, while the former obeys
   $p_i^\mu=m_i u_i^\mu$, where $u_i^2=1$. To achieve definite classical momenta the initial states must be dressed with appropriate coherent\footnote{ Here we think of coherent states in the sense of Perelemov \cite{perelomov:1972}.   We 
   	refer the interested reader to \cite{Perelomov:1986tf} for details of the Perelomov formalism.} wavefunctions, giving us the notion of sharply peaked position and momenta. The most general coherent relativistic wavefunctions associated with the restricted Poincare group have the form \cite{Kaiser:1977ys}
 \begin{align}
 f_z (p):=\braket{e_z|f}=
 \int \dd \Phi(p) e^{-\ii z\cdot p} f(p)  \,,
 \end{align}     
 where $\dd \Phi(p):= (2 \pi)^{d-1}\dd^ d p \delta (p^2-m^2)\Theta(p_0)$ and  
 $\braket {e_z|p}:= \mathcal C_z e^{-\ii p \cdot z}$. Here  $z(t)=x(t)-\ii y$ is a complex vector, which in general is time-dependent. 
 The classical phase space  is obtained by setting $t=0$ in  $x(t)$ so it reduces to a constant \cite{Kowalski:2018xsw}.  The normalization of states can be derived from
\begin{align}
\braket{e_z|e_w}= \int \dd \Phi (p) e^{-\ii p\cdot (z-\bar w)}=
 \mathcal C_ z 
\mathcal C^{*}_w
 \left(\frac{m}{2\pi \eta}\right)^{d/2-1} 
K_{d/2-1}(\eta m),  
\end{align}
where $\eta=\sqrt{-(z-\bar w)^2}$ and  $K_\nu (x)$ is the modified Bessel function.
 For $z=w$,  $\eta=2 |y|$,  one obtains
\begin{align}
\mathcal C_z= \left[{\left(\frac{4 \pi |y|}{m} \right)^{d/2-1}\frac{1}{K_{d/2-1}(2m |y|)}}\right]^{1/2}.
\end{align}
 Wavefunctions employed by KMOC  correspond to the case where one chooses the complex vector to be
\begin{align}
z_i^\mu=-b_i^\mu+\ii \frac{ u_i^\mu}{m \xi}, 
\end{align}
where $\xi$ is a dimensionless parameter, which can be thought as the square of the ratio of the Compton wavelength to the intrinsic spread of the wavepacket.  Here $u_i$ is the classical four velocity of the particle of mass $m_i$.
Therefore,  it is natural to consider the current\footnote{We will keep
	the indices explicit for calculations but otherwise suppress them to avoid cluttered expressions.} weighted with coherent wavefunctions as 
\begin{align}
\label{current1}
&\mathcal{C}_n (p, k):=\int \dd \Phi (p_1, p_2) \phi_{z_1}(p_1) \phi_{z_2} (p_2)  \mathcal{A}_n
(p_1, p_2, k_1, \dots, k_n) \, ,
\end{align} 
where $ \dd \Phi (p_1, p_2):=\dd \Phi(p_1) \dd \Phi(p_2)$. Now the classical limit of the current can
 be computed from 
 a Laurent expansion of the formal expression
\begin{align}
\label{current}
&\mathcal{C}_n (p, k)=\int \dd \Phi (p_1, p_2) \phi (p_1) \phi (p_2) e^{\ii b \cdot \sum_{i=1}^n k_i} \mathcal{A}_n
(p_1, p_2, \hbar k_1, \dots, \hbar k_n) \, ,
\end{align} 
where we have used momentum conservation, $k\to\hbar k$ and $b \to b/\hbar$.  The rescaling has an important effect into the structure of \eqref{current}. Indeed writing explicitly the momentum conservation Dirac-delta  we have
\begin{align}
&\mathcal{C}_n (p, k)= 
\int \dd \Phi (p_1, p_2)  \phi (p_1) \phi(p_2) e^{\ii b \cdot \sum_{i=1}^n k_i} \hdelta^4(p_1+p_2+\hbar \sum_{i=1}^nk_i ) { A}_{n} (p,\hbar k)\\ 
& = 
\int \dd \Phi(p_1) \ \phi (p_1)
\phi(-p_1 - \hbar \sum_{i=1}^n k_i )
\hdelta( \hbar^{2} \sum_{i,j=1}^{n} k_{i}\cdot k_{j} +2 \hbar p_1 \cdot \sum_{i=1}^n k_i ) e^{\ii  b \cdot \sum_{i=1}^n k_i} {A}_{n}(p,\hbar k), \nonumber
\end{align}
where we have used the momentum conservation Dirac-delta to perform the phase-space integral over $p_2$.
Making the identifications $q \to \sum_{i=1}^n k_i$ the remaining integral has the form
\begin{align}
\int  \dd \Phi(p) \phi (p)
\phi (-p- \hbar  q) \hdelta(2\hbar p \cdot q+\hbar^2 q^2) f(p, q),
\end{align} 
which is sharply peaked around $p^\mu=m u^\mu$, where $u^\mu$ is the classical velocity of the particle with mass $m$. The analysis of the above integral by KMOC (see Appendix B of Ref.\cite{Kosower:2018adc}) does not depend on the on-shell properties of 
$q$, which plays the role of the momentum mismatch in KMOC,  so we can apply it here as well. The current then simplifies to
\begin{align}
\mathcal{C}_n (p,  k)= \frac{1}{2}\hdelta\left( p \cdot \sum_{i=1}^n k_i\right)
e^{\ii b \cdot \sum_{i=1}^n k_i}
\bar { A}_{n} (p,k) \,, 
\label{final-current-KMOC}
\end{align}
where $\bar A(p,k)$ denotes the non-vanishing term in the Laurent expansion, where by abuse of notation we have set $p_{1}=p$.  From a practical point of view we are done.  We can already perform calculations following the KMOC algorithm and reach   Eq.\eqref{final-current-KMOC} for the theory under study.

A few comments are in order. Strictly speaking  Eq.\eqref{final-current-KMOC} should be understood as the average of the RHS
over wavefunction of $p$, which in KMOC is denoted by a double bracket. 
 The net effect of weighting over coherent wavefunctions is producing an overall factor depending on the external soft momenta, which is analogous to the momentum mismatch in classical observables.  
 The attentive reader might ask about the presence of singular terms produced by the series expansion.
The current is not an observable so one might expect such terms. However, the current is an off-shell tree-level object so we may safely ignore Feynman's $\ii \epsilon$ prescription in calculations, thus leading to cancellation of those singular terms. In QED we have checked this up to 7-points. In the worldline formulation of the
current the absence of those singular terms will become clear as we discuss now.
  
\subsection{Relation to WQFT }\label{sec-proof-WQFT}

We wish to relate the classical current \eqref{final-current-KMOC} to a worldline path integral. Although we will consider scalar QED as an illustration,  the following discussion is also applicable to other  theories where the WQFT is known. The field theory 
action for scalar electrodynamics 
\begin{align}
S_{\text{sQED}}[\varphi,\varphi^{\dagger},A] =\int \dd^4 x\left[(D^\mu\varphi)^\dagger D_\mu \varphi -m^2 \varphi^\dagger \varphi\right] \, 
\end{align}
with the gauge covariant derivative $D_\mu=\partial_\mu-\ii e A_\mu$. The gauge-fixed Maxwell action is
\begin{align}
S_{\text{gf}}[A]=- \int \dd^4 x \left[\frac 1 4 F_{\mu\nu}F^{\mu\nu}+\frac 1 {2\xi } (\partial^\mu A_\mu)^2\right],
\end{align}
where we will use Feynman gauge ($\xi=1$) in calculations. We know  from standard QFT that the current \eqref{general-def} can be written off-shell as a Fourier transform of a time ordered correlation function of the quantum fields describing the external states after LSZ reduction, which  can be represented as a path integral. Moreover,  we are interested in the classical limit so it will be convenient to perform the LSZ reduction in two steps, first on the photon legs and then on the scalar ones. 
 In momentum space the amputation of the photon external legs leads to
\begin{align} \label{P-off}
{\cal P}_n^{\text{sQED}}(p,k )
=\frac{1}{\cal N}\int \D A \ e^{\ii S_{\text{gf}}[A]} \int& \D\varphi \D \varphi^{\dagger } e^{\ii S_{\text{sQED}}[\varphi,\varphi^{\dagger},A]}\\
& \ii k_1^2 \cdots \ii k_n^2\ \varphi(p_{1})\varphi^{\dagger}(p_{2}) A_{\mu_{1}}(k_{1})\cdots A_{\mu_{n}}(k_{n})
\Big|_{k^2\ne 0} \nonumber
 \,.
\end{align}

Before LSZ let us consider  the classical limit of Eq.\eqref{P-off} following Ref.~\cite{Mogull:2020sak}. In the classical approximation we can neglect loops mediated by scalars and replace the path integral over scalar fields  by the photon-dressed scalar propagator, which we briefly review now\footnote{Dressed propagators have been developed
	in a  worldline representation for a variety of models, see e.g. \cite{Ahmadiniaz:2015kfq,
		Ahmadiniaz:2015xoa, Edwards:2017bte, Ahmadiniaz:2017rrk, Ahmadiniaz:2020wlm, Corradini:2020prz,
		Ahmadiniaz:2021gsd}. A standard reference for worldline methods is Ref.\cite{Schubert:2001he}.}.
	In position space the photon-dressed scalar propagator reads
\begin{equation}
\mathcal G(x_1,x_2)[A] = \frac{1}{{\bar{ N}}}\int \D \varphi \D\varphi^{\dagger} e^{\ii S_{\text{sQED}}[\varphi,\varphi^{\dagger},A]}
\varphi(x_{1})\varphi^{\dagger}(x_{2})\,.
\end{equation}
The dressed propagator admits  a worldline path integral representation of the form
\begin{align}
\mathcal G(x_1,x_2)[A] &= \langle x_{2}|\frac{1}{D^{2} +m^{2} - \ii\epsilon }| x_{1}\rangle \\\nonumber
&= \ii \int_{0}^{\infty}\dd T \,\langle x_{2}| e^{-\ii T (D^{2} +m^{2} + \ii\epsilon )}| x_{1}\rangle
= \int_{0}^{\infty } \dd T \int_{x(0)=x_{1}}^{x(1)=x_{2}} \D x \, e^{-\ii \int_{0}^{1}d\tau \left( \frac{1}{2T}\dot{x}^{2} +e \dot{x}_{\mu}A^{\mu}\right)}\,,
\end{align} 
where $T$ is the so-called Schwinger proper time while the Feynman $\ii\epsilon$ prescription is understood in the path integral. The dressed propagator in momentum space can be written as \cite{Daikouji:1995dz}
\begin{align}
\mathcal G(p,k)[A]=
\hat{\delta}^4 \left(p_1+p_2+\sum_{i=1}^n k_i\right) G (p,k)[A]\,,
\label{dressed-gen}
\end{align} 
whose explicit form is not required for our purposes. Hence the path integral with no matter loops is
\begin{align} \label{P-off2}
{\cal P}_n^{\text{sQED}}(p,k )
=&\frac{1}{\tilde {\cal N}}
\hat{\delta}^4 \left(p_1+p_2+\sum_{i=1}^n k_i\right)\\
&
\int \D A \ e^{\ii S_{\text{gf}}[A]}   G (p,k)[A]\ \ii k_1^2 \cdots \ii k_n^2\ A_{\mu_{1}}(k_{1})\cdots A_{\mu_{n}}(k_{n})\Big|_{k^2\ne 0}^{\text{trees}} \nonumber
\,,
\end{align}
where we have absorbed factors related to our definition of dressed propagator into  $\tilde {\mathcal N}$. This is not yet the classical limit of 
the current we are looking for because we have not performed the LSZ reduction on the external scalar legs, namely 	\cite{Fabbrichesi:1993kz}
 \begin{align}
 G^{\text{c}}(p)[A]= \lim\limits_{p_1^2, p_2^2 \to m^2} \ii(p_1^2-m^2) \ \ii (p_2^2-m^2) \int \dd^4 x \dd^4y \
 e^{\ii p_1\cdot x+\ii p_2\cdot y}G(x, y)[A] \;.
 \label{Gconnected}
 \end{align}
  At the level of the worldline integral we can also achieve the LSZ reduction by
 Fradkin's prescription of exchanging the limit of integration in the worldline action to $(-\infty, +\infty)$ 
 as a consequence of performing the Schwinger proper time integration after amputating the external scalar propagators. 
 The latter step fixes the boundary conditions needed to perform the perturbative expansion of the amputated dressed propagator. The equivalence between both procedures  is encoded in the relation
 \begin{align} \frac{\Sigma(b,p;A)}{\Sigma_{0}}=
 e^{\ii b \cdot \sum_{i=1}^n k_i} \deltahat\left( p\cdot \sum_{i=1}^{n}k_{i}	\right)G^c(p;A) \,, 
 \label{factors-MPS}
 \end{align}
 which was 
 explicitly demonstrated for the graviton-dressed scalar propagator in Ref.\cite{Mogull:2020sak}. Here $\Sigma_{0}$ is some overall factor that we can absorb into the normalization of the correlation function. Notice that both sides depend only on $p_1=p$ on the support of the Dirac-delta in Eq.\eqref{dressed-gen}. The left hand side of Eq.\eqref{factors-MPS} is given by
 \begin{align}
 \Sigma(b,p;A) = \int \D x \exp \left[ 
 -\ii \int_{-\infty}^{\infty}d\tau \left( \frac{1}{2}\dot{x}^{2} + e \dot{x}_{\mu}A^{\mu}(x(\tau))				\right)\right], 
 \label{sigma-path}
 \end{align} 
where  $b$ and $p$ arise
 from  the background expansion $x^{\mu}(\tau)= b^{\mu} + p^{\mu}\tau +q^{\mu}(\tau)$. We will see in later Sections how to evaluate \eqref{sigma-path} from Worldline Feynman Rules (WFRs). 
 
 Going back to Eq.\eqref{Gconnected} and restricting the calculation to tree-level, the current can be written as
 \begin{align} \label{P-off3}
 {\cal A}_n^{\text{sQED}}(p,k )
 =&\frac{1}{\tilde {\cal N}}
 \hat{\delta}^4 \left(p_1+p_2+\sum_i k_i\right)\\
 &
 \int \D A \ e^{\ii S_{\text{gf}}[A]}   G^c(p,k)[A]\ \ii k_1^2 \cdots \ii k_n^2\ A_{\mu_{1}}(k_{1})\cdots A_{\mu_{n}}(k_{n})
 \Big|_{k^2\ne 0, p_i^2 =m^2}^{\text{trees}} \nonumber
 \,,
 \end{align} 
 which brings us closer to the WQFT representation of $\mathcal C^{\text{sQED}}_n (p,k)$. Recalling  Eq.\eqref{general-def} and imposing momentum conservation we have
\begin{align}
\bar A(p,k)=\frac{1}{\tilde {\cal N}}
\int \D A \ e^{\ii S_{\text{gf}}[A]}   G^c(p,k)[A]\ \ii k_1^2 \cdots \ii k_n^2\ A_{\mu_{1}}(k_{1})\cdots A_{\mu_{n}}(k_{n})\Big|_{p_2=-p_1-\sum_i k_i} \, ,
\end{align}
which is a purely tree-level object in the classical approximation which we can identify 
with  $\bar A(p,k)$ in
Eq.\eqref{final-current-KMOC}. Thus, inserting this equation into Eq.\eqref{final-current-KMOC} we have
\begin{align}
\mathcal{C}^{\text{sQED}}_n (p,  k)=&
\frac{1}{\tilde {\cal N}}
e^{\ii b \cdot \sum_{i=1}^n k_i} \deltahat\left( p\cdot \sum_{i=1}^{n}k_{i}	\right)\\
&\int \D A \ e^{\ii S_{\text{gf}}[A]}   G^c(p,k)[A]\ \ii k_1^2 \cdots \ii k_n^2\ A_{\mu_{1}}(k_{1})\cdots A_{\mu_{n}}(k_{n})\Big|_{p_2=-p_1-\sum_i k_i} \, .\nonumber
\end{align}
Finally, applying the relation \eqref{sigma-path} 
we arrive at the WQFT representation of the off-shell current
\begin{align}
\mathcal{C}^{\text{sQED}}_n (p,  k)=
\frac{1}{\cal Z} 
\int \D A \, e^{\ii S_{\text{gf}}[A]} \,
\int \D x e^{
	-\ii \int_{-\infty}^{\infty}d\tau \left[ \frac{1}{2}\dot{x}^{2} + e \dot{x}_{\mu}A^{\mu}(x(\tau))				\right]}   \, \ii k_1^2 A_{\mu_{1}}(k_{1})\cdots \ii k_n^2A_{\mu_{n}}(k_{n}) ,
\label{path-integral-exp}
\end{align}
which generalizes WQFT for an arbitrary number of off-shell photons\footnote{Another alternative to generate photon insertions is to consider the functional with a factor $\exp(JA)$ and take derivatives over $J$.}.  
The factor
${\cal Z}$ ensures that the current is normalized to one when there are  no gauge fields in the path integral and defines the WQFT partition function
\begin{equation}\label{zqed}
{\cal Z} = \int \D A \, e^{\ii S_{\text{gf}}[A]} \, \int \D x \, e^{-\ii \int_{-\infty}^{\infty} \dd\tau \left(	\frac{1}{2}\dot{x}^{2} + e \dot{x}\cdot A	\right) } \, .
\end{equation}
The  boundary conditions on the worldline variables depend on the model under consideration.  Collecting all of the integration constants into $\mathcal Z$ the classical off-shell current in WQFT can be succinctly written as
\ba 
\mathcal{C}^{\text{sQED}}_n (p, k)&= \ii k_1^2 \cdots \ii k_n^2\left \langle A_{\mu_{1}}(k_{1})\hdots A_{\mu_{n}}(k_{n})\right\rangle_{\text{WQFT}}\, .
\label{current-classical-worldlined-QED}
\ea 
Calculations based on Eq.\eqref{path-integral-exp} will produce precisely the factors in  Eq.\eqref{final-current-KMOC} from which we can
extract $\bar A_n(p,k)$. In doing so one must keep in mind the overall factor of two in Eq.\eqref{final-current-KMOC}. Moreover, the presence of the Dirac-delta is required to match currents in both approaches but can be dropped at the end of the computation.   The worldline path integral \eqref{current-classical-worldlined-QED} can be treated perturbatively deriving Feynman rules from the worldline action leading directly to $\bar A_n(p,k)$ without any Laurent expansion. 

Other theories can be treated along the same lines. For gravity we will consider this in Section \ref{computational-part}. 
Gauge theories with color can  be treated along the same lines with the usual caveats surrounding properties of color charges in the classical limit. On the amplitudes side one can follow
Ref.\cite{delaCruz:2020bbn} while in the WQFT front one can apply the methods of Ref.\cite{Shi:2021qsb}.

Let us end this Section by summarizing its key contribution. Starting with an off-shell current we have shown that dressing it with coherent wave-functions \eqref{current}  leads to a WQFT path integral thus finding a direct connection of the KMOC approach and WQFT for off-shell tree-level currents.

\section{Computing off-shell currents}
\label{computational-part}
The evaluation of Eq.\eqref{current-classical-worldlined-QED} can be done by deriving worldline Feynman rules (WFRs), which take care of the perturbative evaluation of the Gaussian path integral. There are two types of Wick contractions: those related with the worldline path integral and  Wick contractions among gauge fields.
The net effect of the contraction among fields will be to generate  permutations. 

The partition functions considered in the rest of the paper have the schematic form
 \begin{align}
 \mathcal Z = \int {\mathcal D} B\, e^{\ii S[B]}
 \int {\mathcal D} x \int \mathcal D \chi e^{\ii S\left[x, \chi; B (x)\right]}\, ,
 \end{align}
 where $B$ represents a background field and  $\chi$ are auxiliary worldline variables. The action $S[B]$ is the gauge fixed action of the background field and   
 $S\left[x, \chi; B (x)\right]$ is the worldline action in WQFT, from which we can derive WFRs. 
 We rescale the worldline parameter  to derive WFRs written in terms of the worldline momentum $p^{\mu}$ instead of $u^\mu$.   The relevant  wordline actions are
 \begin{align}
 S_{\text{BA}}[x,\lambda,\bar{\lambda},\gamma,\bar{\gamma}; \Phi]=&- \int_{-\infty}^\infty \dd \sigma\left(\frac 1 2  \dot x ^2+\ii \bar \lambda_a \dot \lambda^a +\ii \bar \gamma_\alpha \dot \gamma^\alpha - \frac {y} 2  Q^a\Phi^{a\alpha} \tilde Q^\alpha \right)\, , \label{BAction}\\
 S_{\text{YM}}[x, \lambda; A] =& -\int_{-\infty}^{\infty}\dd\sigma\left(
 \frac{1}{2}\dot{x}^{2}+\ii \bar{\lambda}_{a}\dot{\lambda}^{a}+g \dot{x}^{\mu}A_{\mu}^{a}Q^{a} \right)\,, \label{YMC}\\
	S_{\text{GR}}[x, \psi, \bar \psi;A]=& -\int_{-\infty}^{\infty} \dd\sigma \left( \frac{1}{2}g_{\mu\nu}\dot{x}^{\mu}\dot{x}^{\nu} + \ii \bar{\psi}_{a} \frac{D \psi^{a}}{D\tau} - \frac{1}{8}R_{abcd}S^{ab}S^{cd}				\right)\, ,\label{GRSpin}
\end{align}
for Bi-Adjoints, Yang-Mills, and GR, respectively.
The covariant derivative is $\frac{D\psi^{a}}{D\tau} = \dot{\psi}^{a} + \dot{x}^{\mu}\omega_{\mu a b}\psi^{b}$ and $\omega_{\mu a b}= e_{a\nu}(\partial_{\mu}e^{\nu}_{b} + \Gamma^{\nu}_{\mu\lambda} e^{\lambda}_{b} )$ is the spin connection. The translation between curved and flat indices is done by using the tetrad fields $e^{\mu}_{a}$.
The spin tensor of the particle is defined by $S^{\mu\nu} = -2\ii e^{\mu}_{a} e^{\nu}_{b}\ \bar{\psi}^{[a}\psi^{b]}$.
The derivation of WFRs is now well established and described at length
in Refs. \cite{Mogull:2020sak, Jakobsen:2021zvh, Shi:2021qsb, Bastianelli:2021nbs, Wang:2022ntx} so here we only  consider briefly EM. Our treatment of color variables is equivalent to Ref.\cite{Shi:2021qsb} but differs on the introduction of auxiliary color variables\footnote{Color charges are given by $Q^a=\bar \lambda \cdot T^a \cdot \lambda$ and $\tilde Q^\alpha= \bar\gamma \cdot \tilde{T}^\alpha \cdot \gamma$, where $\lambda$ and $\gamma$ are the auxiliary variables. 
	A pedagogical description of auxiliary variables is in Ref.\cite{Bastianelli:2021rbt}. See also Refs.\cite{Edwards:2019eby,Bastianelli:2013pta,Corradini:2016czo}.}.
  The interested reader can see the details of the derivation of WFRs with color in Appendices \ref{bi-adjoint-WQFT}  and \ref{colors-appendix}.

  \subsection{Gauge} 
 \label{gauge-examples}
WFRs are easy to derive  for the colorless version of 
\eqref{YMC}. 
  Inserting the background expansion  $x^\mu (\tau)= b^{\mu} + p^{\mu}\tau + q^{\mu}(\tau)$ into the action
\begin{align}
S_{\text{sQED}}[x; A] =& -\int_{-\infty}^{\infty}\dd\sigma\left(
\frac{1}{2}\dot{x}^{2}+e \dot{x}^{\mu}A_{\mu} 
\right)\,,
\end{align}
with the momentum space expansions
\begin{align}
q^{\mu}(\tau) = \int_{-\infty}^{\infty}\hdd\omega \, e^{\ii \omega \tau}q^{\mu}(-\omega), \  \hskip.5cm A^\mu (x) =  \int_{-\infty}^{\infty}\hdd^4 k \, e^{\ii k \cdot x }
A^\mu(-k),
\end{align} 
  the propagators of the worldline and the gauge field (in Feynman gauge) are
\begin{align} \label{wprop}
\raisebox{-1.7mm}{
	\begin{tikzpicture}[thick]
	\coordinate (A) at (-1,-0);
	\coordinate (B) at (1,-0);
	\filldraw (A) circle (2pt) node[left] {$q^\mu$};
	\filldraw (B) circle (2pt) node[right] {$q^\nu$};
	\draw[] (0,.5) node[]{$\omega$};
	\path[ very thick,draw] (-1,0)--(1,0); 	\path[ thick,draw,->] (-0.5,0.3)--(0.5,0.3) node[pos=.5] {}; 
	\end{tikzpicture}
}
= - \ii \frac{\eta^{\mu\nu}}{(\omega+\ii\epsilon)^2},
\hskip.4cm 
\raisebox{-1.7mm}{
	\begin{tikzpicture}[thick]
	\coordinate (A) at (-1,-0);
	\coordinate (B) at (1,-0);
	\filldraw (A) circle (2pt) node[left] {$A^\mu$};
	\filldraw (B) circle (2pt) node[right] {$A^\nu$};
	\draw[] (0,.6) node[]{$k$};
	\path[ thick,draw,->] (-0.5,0.3)--(0.5,0.3) node[pos=.5] {}; 
	\path[snake it,draw] (-1,0)--(1,0); 
	\end{tikzpicture}
}
= - \ii \frac{\eta^{\mu\nu}}{(k^{2}+\ii\epsilon)} \, . 
\end{align}
  In addition,  the interaction vertices propagating $n$-quantum fluctuations of the worldline variables can be compactly written as (see also Ref.\cite{Wang:2022ntx})
\begin{equation}
\raisebox{-10mm}{\vnQED } = e\,\ii^{n-1} e^{\ii b\cdot k}\deltahat\left(k\cdot p +\sum_{i=1}^{n} \omega_{i}\right) \left(
p^{\mu}\prod_{i=1}^{n}k^{\alpha_{i}} + \sum_{i=1}^{n}\omega_{i}\eta^{\mu \alpha_{i}}\prod_{j\neq i}^{n}k^{\alpha_{j}}
\right)\,, 
\end{equation}
where the solid lines are outgoing. Contractions of gauge fields generate topologies of worldline diagrams, in which external propagators must be amputated in the sense of LSZ. For instance at lower orders  it is easy to see that the Wick contractions among gauge fields lead us to evaluate the diagrams shown in Fig.\ref{main-topologies-off-shell}.

\subsubsection*{Examples}

 \begin{figure}
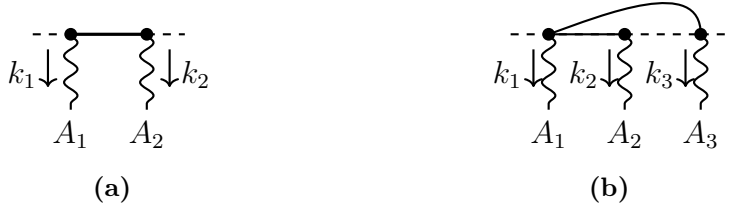

	\centering
	\begin{subfigure}[t]{0.3\textwidth}
		\centering	\lotop{$A_1$}{$A_2$}{black}
		\caption{}
		\label{lo-gen}
	\end{subfigure}
	\begin{subfigure}[t]{0.5\textwidth}
		\centering
		\nlotop{$A_1$}{$A_2$}{$A_3$}{black}{black}
		\caption{}
		\label{nlo-gen}
	\end{subfigure}
	\caption{Worldline diagrams required for the calculation of 2-point (a) and 3-point (b) currents. The dashed lines are drawn only for illustration and represent the worldline. Solid lines represent fluctuations of matter lines and later also auxiliary variables. Wavy lines represent off-shell massless particles and $A$ its associated field. }
	\label{main-topologies-off-shell}
\end{figure}
\begin{figure}[h]
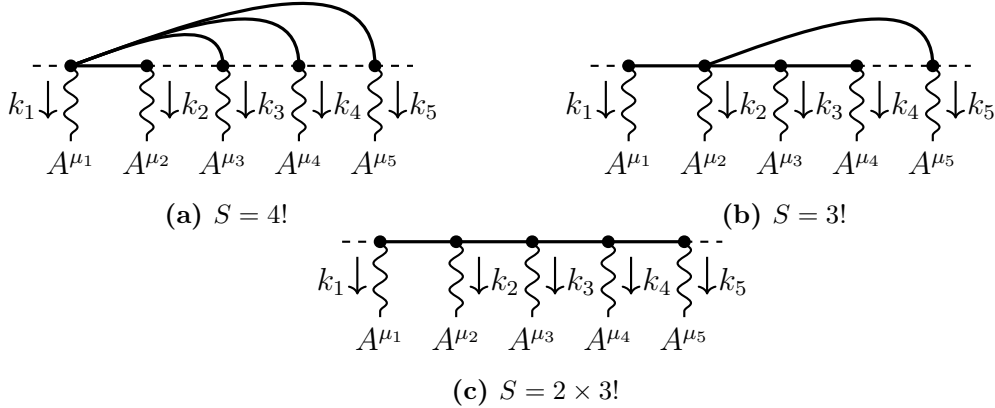

	\centering
	\begin{subfigure}[t]{0.4\textwidth}
		\centering	\lotopfivePHone{$A^{\mu_1}$}{$A^{\mu_2}$}{$A^{\mu_3}$}{$A^{\mu_4}$}{$A^{\mu_5}$}
		\caption{$S=4!$}
	\end{subfigure}
	\begin{subfigure}[t]{0.5\textwidth}
		\centering	\lotopfivePHtwo{$A^{\mu_1}$}{$A^{\mu_2}$}{$A^{\mu_3}$}{$A^{\mu_4}$}{$A^{\mu_5}$}
		\caption{$S= 3!$}
	\end{subfigure}\\
	\begin{subfigure}[t]{0.5\textwidth}
		\centering	\lotopfivePHthree{$A^{\mu_1}$}{$A^{\mu_2}$}{$A^{\mu_3}$}{$A^{\mu_4}$}{$A^{\mu_5}$}
		\caption{$S=2 \times 3!$}
	\end{subfigure}
	\caption{Examples of worldline topologies required for the calculation of 7-point sQED current and their symmetry factors $S$. }
	\label{topologies-five-photons}
\end{figure}

Let us consider the  $n=2$ off-shell current for scalar electrodynamics.
Two equivalent diagrams with symmetry factor $\frac 12$ are generated by the wordline path integration. Hence it is enough to consider the one in Fig.\ref{lo-gen}  whose symmetry factor is unity.
An easy calculation gives
\ba 
\mathcal{C}_{\text{sQED}}^{\mu\nu} (p, k) &=e^2 \, e^{\ii b\cdot (k_1+k_2)} 
 \deltahat\left(p\cdot (k_{1}+k_{2})\right) \bar A_{\text{sQED}}^{\mu\nu}(k_{1},k_{2})\\
&=e^2 \, e^{\ii b\cdot (k_1+k_2)}\deltahat\left(p\cdot (k_{1}+k_{2}) \right) \ii \left(
 \eta ^{\mu  \nu }+\frac{k_2^{\mu } p^{\nu }}{p\cdot k_{1} }-\frac{ k_1^{\nu } p^{\mu }}{p\cdot k_{1} }-\frac{ k_1\cdot k_2
   p^{\mu } p^{\nu }}{(p\cdot k_{1})^2}
   \right),
   \label{2ptsQED}
\ea 
which satisfies the Ward identity $k_{i,\mu} \mathcal{C}_{\text{sQED}}^{\mu\nu} (p, k) = 0$ and matches what one calculates using Eq.\eqref{final-current-KMOC}. A  nontrivial example is the $n=5$ case. Some topologies involved are shown in Fig.\ref{topologies-five-photons}. Its Feynman diagrammatic calculation requires 450 diagrams but  the worldline calculation effectively requires only 12 diagrams summed over the permutations of the external photons with no need to perform any Laurent expansion in $\hbar$, so its calculation is simpler in WQFT. The result obtained with WQFT is lengthy\footnote{The full expression for $\bar A_{\text{sQED}}^{\mu_1\dots\mu_5}$ is attached to this submission in FeynCalc notation.}. Schematically it can written as 
 \begin{align}
 \mathcal{C}_{\text{sQED}}^{\mu_1 \dots \mu_5}=e^5\, e^{\ii b\cdot (k_1+\dots +k_5)} \hdelta \left(p\cdot \sum _{i=1}^5 k_i\right) \bar A_{\text{sQED}}^{\mu_1\dots\mu_5}=e^5\, e^{\ii b\cdot (k_1+\dots +k_5)} \hdelta \left(p\cdot \sum _{i=1}^5 k_i\right) \sum_{i=1}^{2451} a_i \mathsf{T}_i^{\mu_1 \dots \mu_5},
 \label{eq-5-pt}
 \end{align}
 where the sum runs over independent tensor structures. We have compared this result against  the Feynman diagram calculation and found agreement.

\subsection{Gravity}
\label{gr-off-shell}
 The worldline QFT allows us to introduce classical spin effects at almost no cost so we will do so. Let us briefly
summarize the WQFT approach of Ref.\cite{Jakobsen:2021zvh}, which employs
 the so called ${\cal N} = 2$ model. This model captures quadratic in spin effects in classical scattering through the introduction of complex Grassman variables $\psi,\bar{\psi}$ allowing to gauge two super-symmetries on the worldline. The relevant actions to construct the partition function are 
 \begin{equation}
 S_{\textrm{EH}}[g] = -\frac {2}{\kappa^{2}}\int \dd^{4}x\, \sqrt{-g}\, R, \quad S_{\text{gf}} = \int \dd^{4}x \left(	\partial^{\nu}h_{\nu\mu}-\frac 1 2 \partial_{\mu} h^{\nu}_{\nu}\right)^{2},
 \end{equation}
where the space time metric is perturbatively expanded as $g_{\mu\nu}= \eta_{\mu\nu}+\kappa h_{\mu\nu}$. The gauge-fixing term
 imposes a weighted version of the de Donder gauge $\partial^{\nu}h_{\nu\mu} = \frac 12 \partial_{\mu} h^{\nu}_{\nu}$, which leads to the graviton
 propagator
 \begin{align}
 \label{eq:propq}
 \raisebox{-2.8mm}{
 	\begin{tikzpicture}[thick]
 	\coordinate (A) at (-1,-0);
 	\coordinate (B) at (1,-0);
 	\filldraw (A) circle (2pt) node[left] {$h_{\mu\nu}$};
 	\filldraw (B) circle (2pt) node[right] {$h_{\rho\sigma}$};
 	\draw[] (0,.6) node[]{$k$};
 	\path[ thick,draw,->] (-0.5,0.3)--(0.5,0.3) node[pos=.5] {}; 
 	\path[double,snake it,draw] (-1,0)--(1,0); 
 	\end{tikzpicture}
 }
 = \frac{\ii}{2k^{2}} \left(	 \eta_{\mu \rho}\eta_{\nu \sigma} +  \eta_{\mu \sigma}\eta_{\rho\nu} -  \eta_{\mu\nu}\eta_{\rho\sigma} 	\right) \;.
 \end{align}
The partition function can be written as follows
\begin{equation} \label{zgr}
{\cal Z} = \int \D h_{\mu\nu}\, e^{\ii (S_{\textrm{EH}} +S_{\textrm{gf}} ) } \int \D x \D \psi \D \bar{\psi} \, e^{\ii S_{\text{GR}}},\end{equation}
with the worldline action given in \eqref{GRSpin}. The worldline path integral measure includes Lee-Yang ghost fields
\begin{equation}
\D x = \int  Dx DaDbDc\, \exp\left(	-\frac{\ii}{2}\int_{-\infty}^{\infty}\dd\sigma\,  g_{\mu\nu}\left(a^{\mu}a^{\nu} + b^{\mu}c^{\nu}		\right)		\right)
\end{equation}
with $a$ bosonic and $b,c $ fermionic, which make the measure translational invariant. For details on the model we refer the reader to \cite{Jakobsen:2021zvh}.
To derive WFRs in such a case one also needs to background expand $\psi^{a}(\tau) = \theta^{a}+ \Psi^{a}(\tau)$ where complex conjugation leads to the expansion for the barred partner. This yields  the classical value of the spin tensor $\s^{ab} = -2\ii \bar{\theta}^{[a}\theta^{b]}$, which is related to the Pauli-Lubanski spin vector $s^{a} = -\frac{1}{2}\epsilon^{abcd}p_{b}\s_{cd}$.  The required WFRs to perform calculations are given in the Appendix \ref{WFR-gr}. Generalizing the results in Sec.\ref{sec-proof-WQFT}, we write the off-shell current in gravity as
\begin{equation}
{\cal C}_n^{\text{GR}}(p,k) = (- \ii)^{n} k_{1}^{2} k_{2}^{2} \cdots  k_{n}^{2}\braket{ h_{\mu_{1} \nu_{1}}(k_{1}) h_{\mu_{2} \nu_{2}}(k_{2}) \cdots h_{\mu_{n} \nu_{n}}(k_{n})
}_{\text{WQFT}}
\label{current-scalar-GR}
\end{equation} 
understood as an expectation value over the partition function \eqref{zgr}. 

As an example we consider the classical on-shell Compton amplitude\footnote{In Ref.\cite{Saketh:2022wap} a direct classical calculation of the Compton amplitude has been done using the worldline effective theory including cubic in spin corrections.}, which describes the scattering of linearized gravitational waves off a black hole. 
\begin{figure}
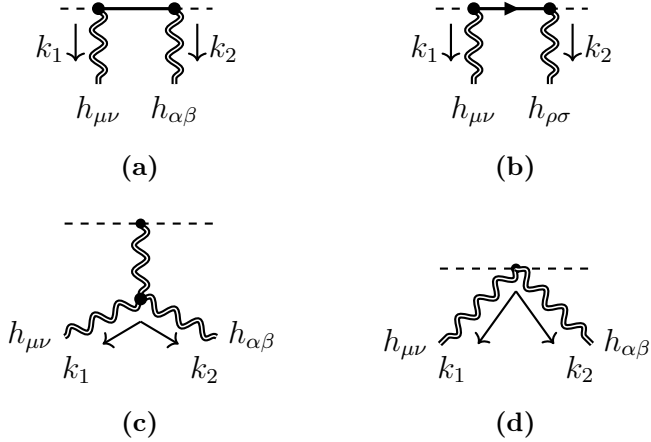

	\centering
	\begin{subfigure}[t]{0.3\textwidth}
			\centering \lotopGR{$h_{\mu\nu}$}{$h_{\alpha\beta}$}{black}
		\caption{}
		\label{tu} 		
		\end{subfigure}
		\begin{subfigure}[t]{0.3\textwidth}
			\centering \lotopGRspin{$h_{\mu\nu}$}{$h_{\rho\sigma}$}{black}
			\caption{}
		\label{tu1} 
		\end{subfigure}\\[1em]
		\begin{subfigure}[t]{.3\textwidth}
		 \centering  \hhh
		 \caption{} 
		 \label{s}
		 \end{subfigure}
		 \begin{subfigure}[t]{.3\textwidth} 
		\centering \hhvertex 
		\caption{}
		\label{seagull} 
	\end{subfigure}
	\caption{Worldline diagrams contributing to the on-shell Compton amplitude. \ref{tu},\ref{tu1}, with crossed topologies are associated with fluctuations  of kinematic and spin variables, respectively.
	 }
	\label{compton-amplitude}
\end{figure}
The on-shell Compton amplitude  is given by
\begin{equation}\label{sGR-Compton}
{\cal M}^{h_{1}h_{2}}(p,k_{1},k_{2}) = 
{\bar M}_{\text{GR}}^{\mu\nu\alpha\beta} (p, k_{1},k_{2}) \epsilon^{h_{1}}_{\mu\nu}(k_{1})\epsilon^{h_{2}}_{\alpha\beta}(k_{2})\Big|_{k_i^2=0}\,,
\end{equation}
where we can extract the amplitude from $\mathcal{C}_{\text{sGR}}^{\mu\nu\alpha\beta}=\frac{1}{2}\kappa^2 e^{\ii b\cdot (k_1+k_2)} \hdelta(p\cdot (k_1+k_2)) \bar M_\text{GR}^{\mu\nu\alpha\beta}$, which receives contributions from the diagrams shown in Fig.\ref{compton-amplitude}. 
In the spinless case, we can apply an off-shell version of the Kawai-Lewellen-Tye \cite{Kawai:1985xq} relation
\begin{align}
{\mathcal C}_{\text{sGR}}^{\mu_1\nu_1,\mu_2\nu_2}=-\kappa^2 \ii \, e^{\ii b\cdot(k_1+k_2)}\hdelta( p \cdot(k_1+k_2)) \frac{k_1\cdot p \, k_2\cdot p}{k_1\cdot k_2} \bar A_{\text{sQED}}^{\mu_1\mu_2}
\bar A_{\text{sQED}}^{\nu_1\nu_2} \,,
\end{align}
with the QED current given in Eq.\eqref{2ptsQED}, after replacing $e^{2}\to \kappa^{2}/4$.
In order to fuse  the full current into the amplitude \eqref{sGR-Compton}, we use physical polarization tensors $\epsilon^{\mu\nu}_{\pm \pm}(k_{i}) = \epsilon^{\mu}_{\pm}(k_{i})\epsilon^{\nu}_{\pm}(k_{i})$ written as a product of null transverse  photon polarizations. We set $k_{1}$ as incoming momentum and $k_{2}$ as outgoing and choose the rest frame of the worldline, i.e.,
\begin{equation}
p^{\mu}=m u^{\mu} = (m,0,0,0),
\quad k^{\mu}_{1} =E(1,0,0,1), \quad
k^{\mu}_{2} = E(1,\sin\theta,0,\cos\theta),
\end{equation} 
 where $E$ is the energy of the graviton.  Explicit polarization vectors $\epsilon^{\mu}_{\pm}$ follow from the transversality and traceless conditions. Therefore, we  can evaluate \eqref{sGR-Compton} for the independent set of helicity configurations $(++),(+-)$. The scalar contributions are given by 
\begin{equation}
|{\cal M}_{++}|_{_{0}} = |{\cal M}_{--}|_{_{0}}= \frac{\kappa ^{2}m^{2}}{4}\frac{\cos^{4}\frac{\theta }{2}}{ \sin^{2}\frac{\theta }{2} },
\hskip.4cm 
|{\cal M}_{+-}|_{_{0}} = |{\cal M}_{-+}|_{_{0}} = \frac{\kappa ^{2} m^{2}}{4} \sin^{2}\frac{\theta }{2}
\,.
\end{equation}
Similarly, using the full current the linear in spin terms read 
\begin{equation}
|{\cal M}_{++}|_{_{O(\s)}} = |{\cal M}_{++}|_{_{0}}\Big|			
s\cdot(k_{1}+k_{2})\tan^{2}\frac{\theta}{2} +\ii \frac{k_{1}\cdot \s\cdot k_{2}}{mE \cos^{2}\frac{\theta}{2}}
\Big|,
\hskip.4cm |{\cal M}_{+-}|_{_{O(\s)}} = |{\cal M}_{+-}|_{_{0}} \, |s\cdot (k_{1}-k_{2})|  \,, 
\end{equation}
while the quadratic in spin contributions can be recast in the suggestive way
\begin{equation}
|{\cal M}_{++}|_{_{O(\s^{2})}} = \frac{1}{2} \frac{\hskip.4cm |{\cal M}_{++}|^{2}_{_{O(\s)}}  }{ |{\cal M}_{++}|_{_{0}}},
\hskip.4cm |{\cal M}_{+-}|_{_{O(\s^{2})}} = \frac{1}{2} \frac{ \hskip.4cm |{\cal M}_{+-}|^{2}_{_{O(\s)}} }{ |{\cal M}_{+-}|_{_{0}}} \,.
\end{equation}
The remaining helicity configurations can be obtained by replacing $\s^{a b }\to -\s^{a b},s^{a}\to -s^{a}$.
The above results are in agreement with \cite{Saketh:2022wap} up to quadratic in spin
upon matching with our conventions.
\section{Application to Hard Thermal loops}
\label{currents-and-HTLs}
\begin{figure}[t]
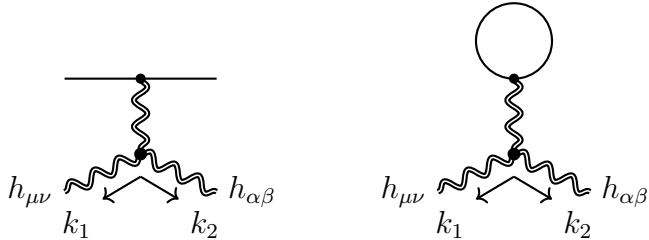

	\centering  \hhhF $ \qquad$
	\hhhFclosed
	\caption{(Left) Example of a Feynman diagram  in the set $\mathcal X$ for the $n=2$ current in GR. (Right)	 The singular diagram produced in the forward limit.} 
	\label{Reg-forward-limit}
\end{figure}

A nice application of the ideas presented in the past sections is the case of Hard Thermal Loops (HTLs). These are  currents in the high temperature limit which can be resumed and incorporated into an effective theory known as HTL effective theory~\cite{Braaten:1989mz,  Frenkel:1989br, Braaten:1990az, Taylor:1990ia,
	Frenkel:1991ts}.  It is well known that the high temperature regime is equivalent to a classical regime. Using a KMOC-like approach this was explicitly demonstrated in  Ref.\cite{delaCruz:2020cpc}, where HTLs were computed  as the limit $\hbar \to 0$. Schematically HTL currents can be written as 
\begin{align} 
\Pi_n(k)=\int \dd\Phi (p) f(p_{0})
 \bar A_n(p, k) \, ,
\label{htl}
\end{align} 
where $f (p_{0})$ is a distribution function at equilibrium and $\bar A_n(p, k)$ is the classical limit of the  current  in the regularized forward limit. The regularization is required  since the same diagrams that contribute to the currents contribute to amplitudes so in general the forward limit is singular.
 Let $\mathcal F$ be the set of all Feynman diagrams contributing to the current \eqref{general-def}. Diagrammatically the regularization consists on dropping the set of all diagrams
producing zero momentum internal edges $\mathcal X$ (see Fig.\ref{Reg-forward-limit}) in the forward limit\footnote{
	These problematic diagrams appear e.g.,  Eq.\eqref{current-classical-worldlined-QED} and \eqref{current-scalar-GR}
	so the regularization should also be implemented  in the worldline formalism.}.  It is defined by
\begin{align}	
A_n (p,k):=\sum_{G\in \mathcal F   \backslash \mathcal X}
d(G)\, ,\label{reg-current}
\end{align}
where $d(G)$ is a rational  expression of the form $N(G)/D(G)$.   In the forward limit $p_1=-p_2$ so momentum conservation becomes
\begin{align}
\sum_{i=1}^{n}k_{i} =0.
\label{momentum-conservation}
\end{align}
The classical limit of Eq.\eqref{reg-current} is obtained through Eq.\eqref{final-current-KMOC}. These currents have been considered in Refs.\cite{delaCruz:2020cpc, delaCruz:2022nlj}.   

The WQFT approach gives us a new way of obtaining these currents. In QED the equivalence between
\eqref{final-current-KMOC} and \eqref{current-classical-worldlined-QED} implies that the  $n-$point HTL can be read off from
\begin{equation}
 \ii \  \hdelta \left(p\cdot \sum\limits_{i=1}^n k_i \right)\frac{1}{2}{\bar A}_n (p, k) = \ii k_{1}^{2}\ii k_{2}^{2}\cdots \ii k_{n}^{2} \langle
A_{\mu_{1}}(k_{1})A_{\mu_{2}}(k_{2})\cdots A_{\mu_{n}}(k_{n}) \rangle_{\text{WQFT}}, 
\end{equation}
where the regularization is  understood in  both sides. Since we are interested in $\bar A_n(p,k)$ we will strip-off the Dirac-delta  produced by WQFT.
Inserting
the RHS of this equation  side into Eq.\eqref{htl} gives and alternative worldline path integral representation of the HTL resumed current.  
A similar matching can be used to obtain HTLs in other theories.

Diagrams that contribute at each order in perturbation can be determined from dimensional analysis.
The treatment of colour in the classical limit requires special care due to  nontrivial cancellations between color and kinematics.
 In the worldline context this is translated into the dynamics of auxiliary variables which are present in the worldline action and give additional interactions. Their role is to encode non commutativity of color factors in general. The fact that classical color factors commute in the classical limit is recovered after integration over  the auxiliary variables, a property that requires representations of the gauge group to be large
  (See Appendix \ref{colors-appendix}). 
 
  When classical color factors are recovered,  thermal currents are obtained after phase-space integration over color.  Phase space integration over classical color factor is defined by
 \begin{align}
 \dd c:=&  \dd^8 c\, c_R \delta(c^a c^b \delta^{ab}-q_2 ) \delta(d^{abc}c^a c^bc^c-q_3 ), \label{DIPSC}
 \end{align}
 where $q_2$ and $q_3$ are Casimir invariants. The factor $c_R$ ensure that the color measure is normalized to unity and we have set the gauge group to be $SU(3)$.  For bi-adjoint scalars we will take two copies of the phase-space integration measure.

\subsection{Scalars}
\label{Bi-adjoint}
As our first application let us consider  scalar theories with cubic interactions. For instance, the bi-adjoint field theory Lagrangian  is given by \cite{Luna:2015paa,White:2016jzc}
\begin{align} 
S_{\text {BA}}[\varphi]=\int\dd^4 x \left[\frac 12 \partial_\mu \varphi^{a \alpha} \partial^\mu \varphi^{a \alpha} 
-\frac{m^2}{2} \varphi^{a \alpha}\varphi^{a \alpha} 
+\frac{y}{3!} f^{abc} \tilde f^{\alpha \beta \gamma}
\varphi^{a \alpha} \varphi^{b \beta} \varphi^{c \gamma}\right] \:, 
\label{lag-phi3}
\end{align}
where $m$ is the mass and $y$ the coupling constant. The bi-adjoint scalar field  $\varphi^{a\alpha}$ transforms under the adjoint representation for each factor of its globally symmetry group $G \times \tilde G$. The Lie algebra for each factor has the form $[T^a, T^b]=f^{abc} T^c$ and the adjoint representation is given by its structure constants, i.e.
$(T_A^a)^{b}_{\ c}=-  f^{abc}$. Throughout we  use Greek indices   for the group $\tilde G$.  We briefly outline the derivation of WFRs for this theory in Appendix \ref{bi-adjoint-WQFT}. It
is instructive to present the colorless and colorful cases separately to see how the appearance of color affect the final results. 
\subsubsection*{Cubic scalars}
Let's consider the simplest example, where we have two soft scalars. Two equivalent diagrams with symmetry factor $\frac 12$ are generated by the path integration over scalar fields. Hence it is enough to consider the one in Fig.\ref{lo-gen}  whose symmetry factor is unity. Using WFRs we obtain\footnote{Notice that the propagators considered for our currents do not have an 
	$\ii \epsilon$ term. In order to recover the retarded temperature dependent currents we consider the analytic continuation $k^0_n \rightarrow 
	k^0_n +\ii \epsilon$
	and $k^0_i \rightarrow k^0_i -\ii \epsilon $, for $i=1, \dots, n-1$,	where we assume that the vertex corresponding to $k_n$ corresponds to the one with the largest time~\cite{Brandt:1998gb}.} 
\begin{align} \label{tp1}
  \raisebox{-.9cm}{
\lotop{$\varphi(k_1)$}{$\varphi(k_2)$}{black}}=
 \ii \frac{1}{2} 
 \bar A_2 (p,k_1)
=& \ii  
\frac{1}{4}y^2 \frac{k_1^2}{(k_1\cdot p)^2}\, ,
\end{align}
which we can use to extract $\bar A_2 (p,k_1)$. 
Hence using momentum conservation, relabeling the independent momentum $k_1=k$ 
we obtain
\begin{align}
\Pi_2 (k)= \frac{y^2}{2} \int\dd \Phi (p) f(p_0)
 \frac{k^2}{(k\cdot p)^2}.
\end{align}
The same procedure can be carried on in order to evaluate the three point thermal current. Notice however that a diagram analogous to Fig.\ref{s} also contributes to the current but is singular in the forward limit so we discard it due to the regularization.  The result now depends on permutations of the 
diagrams in Fig.\ref{nlo-gen}, which can be easily evaluated
\begin{align}
\bar A_3(k_1,k_2,k_3)= \frac{y^3}{4} 
\sum\limits_{\sigma \in \text{Cyclic}} \frac{k_{\sigma_1}\cdot k_{\sigma_2} k_{\sigma_2}\cdot k_{\sigma_3}}{\left(p\cdot
	k_{\sigma_1}\right){}^2 \left(p\cdot k_{\sigma_3}\right){}^2},
\end{align}
where the sum runs only over cyclic permutations of $\{1,2,3\}$. This equation is in agreement with Ref.\cite{delaCruz:2022nlj} obtained from semi-classical kinetic theory.

\subsubsection*{Bi-adjoint}
The bi-adjoint scalar also generates Feynman rules which involve color-color fluctuations and color-matter fluctuations, as explained in Appendix \ref{bi-adjoint-WQFT}. It receives the contributions from the diagrams\footnote{In the colored case, color fluctuations are of the same order in coupling as those with matter. However upon restoring $\hbar$ one must also take into account that color factorization  brings an additional power of $\hbar$. See Appendix \ref{colors-appendix}.} with kinematic fluctuations of the worldlines
 \begin{align}
\raisebox{-0.7cm}{ \lotop{}{}{black} } 
   = \ii \frac{y^2}{4} c^{a_1} c^{a_2} \tilde{c}^{\alpha_1} \tilde{c}^{\alpha_2} \frac{k_1\cdot k_2}{(k_1\cdot p)^2}\,,
   \end{align} 
   and  from the diagrams propagating color fluctuations
   \begin{align}
 \raisebox{-0.7cm}{ \lotopcol{}{}{red}} +
 \raisebox{-0.7cm}{ \lotopcolCross{}{}{red}} 
 =&\, \ii \frac{y^2}{4} \frac{ \tilde{c}^{\alpha_1} \tilde{c}^{\alpha_2}}{k_1\cdot p}\bar{u}\cdot(T^{a_1}\cdot T^{a_2}-T^{a_2} \cdot T^{a_1}  )\cdot u\, , \label{cont2-bi}\\
\raisebox{-0.7cm}{ \lotopcol{}{}{blue}}
+
 \raisebox{-0.7cm}{ \lotopcolCross{}{}{blue}} 
=&\, \ii \frac{y^2}{4}    \frac{c^{a_1} c ^{a_2}}{k_1\cdot p}\bar{v}\cdot( \tilde T^{\alpha_1}\cdot \tilde T^{\alpha_2}-\tilde T^{\alpha_2}\cdot \tilde T^{\alpha_1})\cdot v\label{cont3-bi}\, .
\end{align}
It should be noticed how adding up the two topologies in each of Eq.\eqref{cont2-bi} and \eqref{cont3-bi} generates the structure constants of the Lie algebra 
\begin{align}
\bar{u}\cdot \left( T^{a_1}\cdot T^{a_2}-T^{a_2}T^{a_1} \right)\cdot u = f^{a_1 a_2 a_3}c^{a_3}\, 
\label{Lie-bracket}
\end{align}
where $c^a=\bar u\cdot T^a \cdot u$ and
the same for the tilded partner.
Then, the current simplifies to 
\begin{align}
\bar A_2^{a_1 \alpha_1, a_2 \alpha_2}=  \frac{y^2}{2}\left( c^{a_1} c^{a_2} \ct^{\alpha_1} \ct^{\alpha_2} \frac{ k_1^2}{(k_1 \cdot p)^2}+
\frac{\ct^{\alpha_1} \ct^{\alpha_2}f^{a_1 a_2 a_3} c^{a_3} +c^{a_1} c^{a_2}\tilde f^{\alpha_1 \alpha_2 \alpha_3} \ct^{\alpha_3}}{k_1\cdot p} 
\right)\, .\end{align} 
The phase-space integration over color can be done using the identities
\begin{align}\label{psint}
\int \dd c\ c^{a}=0, \qquad \int \dd c\ c^{a} c^{b}= \delta^{ab},
\end{align}
which follow from Eq.\eqref{DIPSC}. 
Hence after some relabeling
\begin{align}
\Pi^ {a_1 \alpha_1, a_2 \alpha_2}(k)= \delta^{a_1 a_2}\delta^{\alpha_1\alpha_2 } \frac{q_{2}^{2}y^{2}}{2}\int\dd \Phi(p) \frac{ k^2}{(k \cdot p)^2},
 \end{align}
which 
 is in agreement with kinetic theory of Ref.\cite{delaCruz:2022nlj}.
\subsection{Gauge and gravity}
\label{gauge}
As our final example, we 
move directly to scalar QCD since 
 scalar QED HTLs can be recovered 
from the off-shell currents in 
Section \ref{gauge-examples} using momentum conservation. In QED no singular diagrams arise in the forward limit.  
 For instance, the 5-pt HTL is straightforward to obtain from Eq.\eqref{eq-5-pt} and Eq.\eqref{momentum-conservation}. 

Let us consider the 3-point HTL. 
From the WFRs in Appendix 
\ref{colors-appendix} there are two
diagrams we need to calculate. For example, 
for the permutation $\sigma=(1,2,3)$ for the external gluons one has 
\begin{align}
\raisebox{-4mm}{ \nlotopC } &= g^{3}\ \bar u\cdot T^{a_{1}}\cdot T^{a_{3}}u \ c^{a_{2}}
\frac{p^{\mu_{3} }}{p\cdot k_{3} }\bar A_{\text{sQED}}^{\mu_{2}\mu_{1}} (k_{2},k_{1})\,, 
\\
\raisebox{-4mm}{ \nlotopD } &=  -g^{3}\ \bar u\cdot T^{a_{3}} \cdot T^{a_{1}}u\ c^{a_{2}}
\frac{p^{\mu_{3} }}{p\cdot k_{3} }\bar A_{\text{sQED}}^{\mu_{2}\mu_{1}} (k_{2},k_{1})\,,
\end{align}
which  generate the $SU(N)$ structure constants as in Eq.\eqref{Lie-bracket}. Performing the phase space integration \eqref{psint} and summing over all permutations
 the final answer can be written as 
\begin{equation}
\bar A^{\mu_{1} \mu_{2} \mu_{3}}_{a_{1} a_{2} a_{3}}=- 2g^{3}  \sum_{\sigma \in S_3}
f^{a_{\sigma_1 } a_{\sigma_2} a_{\sigma_3}} 
\frac{p^{\mu_{\sigma_3}}}{p\cdot k_{\sigma_3}} \bar A_{\text{sQED}}^{\mu_{\sigma_2}\mu_{\sigma_1}} (k_{\sigma_2},k_{\sigma_1})\label{current-3pt}
\end{equation}
where $S_3$ is the set of all permutations of $\{1,2,3\}$. 
The above result satisfies the identity 
\begin{equation}
 k_{3\mu_{3}} \, {\bar A}^{\mu_{1} \mu_{2} \mu_{3}}_{a_{1} a_{2} a_{3}} = 2g^{3} f^{a_{1} a_{2} a_{3}} \left(
\bar A_{\text{sQED}}^{\mu_{1}\mu_{2}} (k_{1},-k_{1}) -\bar A_{\text{sQED}}^{\mu_{1}\mu_{2}} (k_{2},-k_{2})
\right)
\end{equation}
and can be
straightforwardly brought into the form given in Ref.\cite{delaCruz:2020cpc}. The form of Eq.\eqref{current-3pt} shows the direct connection between QED and QCD in the high temperature regime.

We have checked that the WQFT  approach  reproduces the higher-point 
examples considered in Refs.\cite{delaCruz:2020cpc, delaCruz:2022nlj}
including the  
gravitational case. It is then safe to  conjecture that the remarkable simple expression
\begin{equation}
 k_{1}^{2}k_{2}^{2}\cdots k_{n}^{2} \langle
A^{I_1}(k_{1})A^{I_{2}}(k_{2})\cdots A^{I_{n}}(k_{n}) \rangle_{\text{WQFT}}
\end{equation}
encodes  the resumed HTL expansion for any theory where scalars interact with other scalar,  gauge bosons or gravitons. 

\section{Conclusions}
\label{conclusions}
We have exposed a relation between the  WQFT path integral and the classical limit of an off-shell current. The latter
obtained through dressing up an off-shell current with first quantized coherent states and its classical limit obtained through a KMOC-like
procedure. Using a localized wave-packet to
compute classical  currents is equivalent to pick up the background expansion $x^{\mu} = b^{\mu }+p^{\mu}\tau +q^{\mu}(\tau)$ and setting $\tau\in\, (-\infty,\infty)$ in the worldline action as we have shown in Sec.\ref{off-shell-classical}.
We then constructed  the path integral representation of the Green functions associated with the off-shell current and found a link to the worldline theory in the classical limit, thus 
recasting the current as an expectation value of soft fields inside a worldline path integral. 

We have applied our methods
in various settings. In QED we have tested the validity of our approach up to  7-point, while in gravity we have verified that the 
${\cal N}=2$ susy model correctly reproduces the classical Compton amplitude up to quadrupole finding agreement with  Ref.\cite{Saketh:2022wap}. For colored scalars and QCD, we have applied our methods to obtain HTLs, which are related to the classical limit of off-shell currents \cite{delaCruz:2020cpc}. 
It is straightforward to extend our approach to include more particles either massive or massless. For instance one could easily include photons using the dressed propagator of Ref.\cite{Bastianelli:2021nbs}, while on the KMOC side massless particles  were studied in Ref.\cite{Cristofoli:2021vyo}. 
 
Our work shows how to map computations of the classical
contributions of the off-shell  (amplitude-like) current and the
calculation based 
on worldline path integration, thus providing a route to test future WQFT developments. 
For instance a possible
application of our currents would be the case of  
massive higher-spin particles, where one  could compare the off-shell current in the classical limit and the output produced by WQFT.

Off-shell currents  are usually  computed recursively in high-energy-physics applications  so it would be interesting to  recursively compute currents WQFT too. Hints of their recursive structure have already appeared in Ref.\cite{Jakobsen:2021zvh}. Although we have considered only tree-level currents our construction can be relaxed to include  loops, which can be used in the calculation of loop-level
corrections to the background metric in general relativity  \cite{DOnofrio:2022cvn,Mougiakakos:2020laz,Jakobsen:2020ksu}. On the other hand, cuts that contribute classically 
  might arise from sewing  tree-level currents as it is usual in the generalized unitarity method 
  \cite{Bern:1994cg, Bern:1994zx}. 
   
\addsec{Acknowledgements}
FC would like to thank Canxin Shi for helpful discussions. 
 LDLC acknowledges financial support from the Open Physics Hub at the Physics and Astronomy Department in Bologna. His work is also supported by 
the European Research Council, under grant ERC-AdG-885414. 
 Some of the 
calculations in this paper were done with Feyncalc \cite{Mertig:1990an, Shtabovenko:2016sxi, Shtabovenko:2020gxv}.
\appendix
\section{WFRs for gravity}
\label{WFR-gr}
Starting from the action \eqref{GRSpin}
one also needs to background expand $\psi^{a}(\tau) = \theta^{a}+ \Psi^{a}(\tau)$ and its conjugated in addition to $x^\mu$. This yields to the classical value of the spin tensor $\s^{ab} = -2\ii \bar{\theta}^{[a}\theta^{b]}$, related to the Pauli-Lubanski spin vector  by $s^{a} = -\frac{1}{2}\epsilon^{abcd}p_{b}\s_{cd}$. 
The propagation of fluctuations related to the Grassmann variables (spin variables) implies the use of the following propagator 
\begin{equation}
\raisebox{-1.7mm}{
	\begin{tikzpicture}[thick]
	\coordinate (A) at (-1,-0);
	\coordinate (B) at (1,-0);
	\filldraw[black] (A) circle (2pt) node[left,black] {$\Psi^{\mu}$};
	\filldraw[black] (B) circle (2pt) node[right,black] {$\bar{\Psi}^{\nu}$};
	\draw[] (-0,0.1) node[above] {$\omega$};
	\path[very thick, draw] (-1,0)--(1,0); 
	\path[very thick, draw, ] (-.8,0)--(.1,0)node[currarrow,pos=.9, 
	xscale=1,
	sloped,
	scale=1]{}; 
	\end{tikzpicture}
}
=- \ii \frac{\eta^{\mu\nu}}{\omega+\ii\epsilon}
\end{equation}
alongside with the worldline propagator for the kinematical fluctuations given in Eq.\eqref{wprop}.
For our applications we need the interaction vertices describing the emission of a graviton from the worldline and propagating fluctuations of the configuration space variables. We start by lowest order vertex, namely
\ba \label{frgravity}
\raisebox{-10mm}{\vkzeroGR}&= -\frac{\ii\kappa}{2}\deltahat(k\cdot p)e^{\ii b\cdot k}\left(
p^{\mu}p^{\nu} + \ii (k\cdot \s)^{(\mu} p^{\nu)} -\frac{1}{2} (k\cdot \s)^{\mu}(k\cdot \s)^{\nu}
\right)
\ea
alongside with the vertices propagating kinematical and spin fluctuations
\begin{align}
\raisebox{-10mm}{\vkoneGR } &= \frac{\kappa}{2} \, \deltahat(k\cdot p+\omega )e^{\ii b\cdot k}\\[-1em]
&\Big[
2\omega p^{(\mu}\delta^{\nu)}_{\alpha} + p^{\mu}p^{\nu}k_{\alpha} +\ii (k\cdot \s)^{(\mu}\left(
k_{\alpha }p^{\nu)} + \omega \delta^{\nu)}_{\alpha}
\right)-\frac{1}{2}k_{\alpha}(k\cdot \s)^{\mu}(k\cdot \s)^{\nu}
\Big]\, ,\nonumber\\
\raisebox{-10mm}{\vGRspin} &= -\ii \kappa \deltahat(k\cdot p+\omega)e^{\ii b\cdot k}
\left[
k_{[\rho}\delta_{\sigma]}^{(\mu}(p^{\nu)}+\ii (k\cdot \s)^{\nu)} )
\right]\bar{\theta}^{\sigma}\, 
\end{align}
with the interaction vertex for $\bar{\Psi}$ obtained by reversing the arrow and replacing $\bar\theta \to \theta$.
Finally, we list the vertex describing the emission of two gravitons from the worldline, produced by gauge invariance of the worldline action
\begin{align}
&\raisebox{-10mm}{\hhvertex} \\
&\qquad \qquad = -\frac{\kappa^{2}}{4} \deltahat(p\cdot (k_{1} + k_{2}))e^{\ii b\cdot k}
\Big( (k_{1} \cdot \s)^{\mu_{2}} p^{\mu_{1}}\eta^{\nu_{1} \nu_{2}}-\s^{\mu_{1} \mu_{2}} ( p^{\nu_{1}} k_{1} ^{\nu_{2}} -\frac{1}{2} k_{1}\cdot p\, \eta^{\nu_{1} \nu_{2}})\nonumber\\
&\qquad \qquad +\ii \eta^{\nu_{1} \nu_{2}}( ( k_{1}\cdot \s)^{\mu_{1}}( k_{1}\cdot \s)^{\mu_{2}}+\frac{1}{2} ( k_{2}\cdot \s)^{\mu_{1}} ( k_{1}\cdot \s)^{\mu_{2}} -\frac{1}{2}\s^{\mu_{1} \mu_{2}}(k_{1}\cdot \s\cdot k_{2}))\nonumber\\
&\qquad \qquad +\frac{\ii}{4} k_{1}\cdot k_{2} \s^{\mu_{1}\nu_{2}}\s^{\mu_{2} \nu_{1}} +\ii k_{1}^{\nu_{2}}((k_{1}+k_{2})\cdot \s)^{\mu_{1}} \s^{\mu_{2} \nu_{1}} + 1\leftrightarrow  2 \Big)\;.\nonumber
\end{align}

\section{Bi-adjoint scalar}
\label{bi-adjoint-WQFT}

Going back to the worldline approach let us discuss the bi-adjoint scalar in the classical limit using on-shell dressed propagators.
Following Ref.\cite{Bastianelli:2021rbt} the
action in configuration space reads
\begin{equation} 
S_{\text{BA}}= \theta s+ \phi \tilde s -\int_{0}^{1}d\tau \left(
\frac{1}{2T}\dot{x}^{2} +\ii \bar{\lambda}_{a}\left(	\partial_{\tau} + \ii \theta \right)\lambda^{a} +\ii \bar{\gamma}_{\alpha}\left(	\partial_{\tau} + \ii \tilde{\phi}\right)\gamma^{\alpha} -\frac{y}{2}T Q^{a}\varphi^{a\alpha}\tilde{Q}^{\alpha} 
\right), 
\end{equation} 
where we have chosen the gauge $(e,a,\tilde{a})=(T,\theta,\phi)$ and   $\theta,\phi \in [0,2\pi]$ are interpreted as angles. Further, we have defined the color charges $Q^{\alpha}= \bar{\lambda}\cdot T^{\alpha}\cdot \lambda$ and $\tilde{Q}^{\tilde{\alpha}}= \bar{\gamma}\cdot \tilde{T}^{\tilde{\alpha}}\cdot \gamma$ with the gauge group generators $T,\tilde{T}$ in the adjoint representation.
The scalar-dressed  propagator then reads
\begin{align} 
G[x,u,v,\bar{u},\bar{v}] = \int_{0}^{2\pi}\frac{\dd\theta}{2\pi} \int_{0}^{2\pi}\frac{\dd\phi}{2\pi} \int_{0}^{\infty}\dd T \, e^{-\ii T(p^{2} -m^{2})}\int \D x \, \D \lambda \D \bar{\lambda}\D \gamma \D \bar{\gamma}\, e^{\ii S_{\text{BA}} -\bar{\lambda}\cdot \lambda(1)-\bar{\gamma}\cdot \gamma(1)},
\end{align}
where the boundary conditions 
\begin{align}
(\lambda(0),\gamma(0))=(u,v), \quad  (\bar{\lambda}(1),\bar{\gamma}(1))=(\bar{u},\bar{v}),\quad  (x(0),x(1))=(0,x)
\end{align} 
are consistent with the path integral on the line. Now, given the kinetic terms for the color variables appear with covariant derivatives, we twist them as $\lambda_{a}(\tau)\to e^{-\ii \theta \tau}\lambda(\tau), \gamma_{a}(\tau)\to e^{-\ii \phi \tau}\gamma(\tau) $ to  absorb the connection $\ii\theta$.
Then, we split the color variables as
\begin{align}\label{cd}
\bar \lambda^a(\tau)=& z \bar u^a + \bar \beta^a (\tau), \qquad 
\lambda^a(\tau)= u^a + \beta^a (\tau),\\
\bar \gamma^\alpha(\tau)=& w \bar v^\alpha + \bar \eta^\alpha (\tau), \qquad 
\gamma^\alpha(\tau)= v^\alpha + \eta^\alpha (\tau).
\end{align} 
The color fluctuations $(c,d)$ then acquire Dirichlet boundary conditions i.e. $c(0)=\bar{c}(1)=0$ and similarly for the adjoint partner.
Finally, amputating the external legs and introducing the change of variables  $z=e^{\ii\theta},w=e^{\ii\phi}$ in the angle integration, we can write the partition function for the  bi-adjoint scalar as
\begin{equation}
{\cal Z} = {\mathcal N}_{\text{BA}} \oint \frac{\dd z}{2\pi \ii} \frac{e^{z \bu\cdot u}}{z^{s+1}}\oint \frac{\dd w}{2\pi \ii} \frac{e^{w \bar{v}\cdot v}}{v^{\tilde{s}+1}} \, \int \D \varphi \, e^{\ii S[\varphi]}\int  \D \mathsf X\, e^{\ii S[\mathsf X;\varphi]} \,,
\label{functional}
\end{equation}
with $\mathsf X = (x,\beta,\bar{\beta},\eta,\bar{\eta})$, ${\mathcal N}_{BA}$ a normalization constant to be fixed later on, while the WQFT action reads as
\begin{equation}
S[\mathsf X,\varphi] = -\int_{-\infty}^{\infty}d\sigma \left( \frac{1}{2}\dot{x}^{2} +\bar{\beta}_{a}\beta^{a} + \bar{\eta}_{\alpha}\dot{\eta}^{\alpha}-\frac{y}{2} Q^{a}\varphi^{a\alpha}\tilde{Q}^{\alpha} \right)
\end{equation}
with the background expanded color charges written as
\begin{align}
Q^{a}= \left(z \bar{u} + \bar{\beta}  \right)\cdot(T^{a})\cdot \left(u+\beta\right),\quad  \tilde{Q}^{\alpha}= \left(w \bar{v} + \bar{\eta}  \right)\cdot(\tilde{T}^{\alpha})\cdot \left(v+\eta\right) \,,
\end{align}
which allows us to identify the classical color charges as $c^{a}= \bar{u}\cdot T^{a}\cdot u$ and $\tilde{c}^{\alpha}= \bar{v}\cdot \tilde{T}^{\alpha}\cdot v$ up to a normalization factors, as we will see below.
Notice that the accompanying factors of the contour integrals in Eq.\eqref{functional} arise from boundary terms due to the splitting the color variables. 
Let us now move to the interaction vertices.
Since we already background expanded the color variables, we just need to expand the configuration space variables as $x^{\mu}(\tau) = b+p^{\mu}\tau+q^{\mu}(\tau)$.  In momentum space,  the quantum fluctuation of the color variable
is
\begin{equation} 
\beta_{a}(\tau) = \int_{-\infty}^{\infty}\hat{\dd}\omega \, e^{\ii \omega \tau} \beta_{a}(-\omega)\, ,
\end{equation}
and a similar expression for the  color variable $\eta(\tau)$.
This allows first to derive the worldline propagator for  kinematical and color fluctuations as
\begin{align}
\raisebox{-1.7mm}{
	\begin{tikzpicture}[thick]
	\coordinate (A) at (-1,-0);
	\coordinate (B) at (1,-0);
	\filldraw (A) circle (2pt) node[left] {$q^\mu$};
	\filldraw (B) circle (2pt) node[right] {$q^\nu$};
	\draw[] (0,.5) node[]{$\omega$};
	\path[ very thick,draw] (-1,0)--(1,0); 	\path[ thick,draw,->] (-0.5,0.3)--(0.5,0.3) node[pos=.5] {}; 
	\end{tikzpicture}
}
= - \ii \frac{\eta^{\mu\nu}}{(\omega+\ii\epsilon)^2}\,,\hskip.3cm  
\raisebox{-1.7mm}{
	\begin{tikzpicture}[thick]
	\coordinate (A) at (-1,-0);
	\coordinate (B) at (1,-0);
	\filldraw[red] (A) circle (2pt) node[left,black] {$\beta^{\mu}$};
	\filldraw[red] (B) circle (2pt) node[right,black] {$\bar{\beta}_{\nu}$};
	\draw[] (-0,0.1) node[above,] {$\omega$};
	\path[](-1,0)--(1,0) node[color=red,currarrow,pos=.5, 
	xscale=1,
	sloped,
	scale=1]{};
	\path[very thick, red,draw] (-1,0)--(1,0); 
	\end{tikzpicture}
}
=- \ii \frac{\delta_\nu\,^{\mu}}{\omega+\ii\epsilon}
\end{align} 
with the adjoint fluctuations $\eta(\omega)$ having the same propagator as their partner.
We start by writing down the general $n-$point vertex propagating only quantum fluctuations of the configuration space variables  
\begin{align}
\raisebox{-15mm}{
	\begin{tikzpicture}[thick]
	\path [dashed, draw]
	(-1,-1) -- (0,-1);
	\draw[very thick] (0,-1) arc (-180:-280:1)node[right]
	{$q_{\mu_n}(\omega_{n})$};
	\draw[very thick] (0,-1) arc (-210:-280:1)node[right]
	{$q_{\mu_2}(\omega_{2})$};
	\path [dashed, very thick, dots=3 per .4cm, draw](.8,-.4) -- (.8,-0.1);
	\path [very thick,draw]
	(0,-1) -- (1,-1)node[right]
	{$q_{\mu_1}(\omega_{1})$};
	\path [draw,snake it]
	(0,-1) -- (0,-2)node[below]{$\varphi^{a\alpha}(k)$};
	\draw[->,>=stealth] (-0.5,-1.2) -- (-0.5,-1.8) node[midway, left]
	{$k$};
	\coordinate (A) at (0,-1);
	\filldraw (A) circle (1.5pt);
	\end{tikzpicture}
}=\frac y 2 c^{a} \, \tilde{c}^{\alpha} (\ii)^{n+1} e^{\ii b\cdot k} \deltahat\left(	k\cdot p +\sum_{j=1}^n \omega_j	\right)\prod_{i=1}^{n}k^{\mu_i} \, ,
\label{n-point-cubic}
\end{align}
which is the main ingredient needed to perform calculations.
Next, the introduction of color generates WFRs propagating fluctuations of the auxiliary bosonic variables
\ba
\raisebox{-10mm}{\vrightAdj} &= \ii  \frac y 2 \tilde{c}^{\alpha}e^{\ii b\cdot k} \, \deltahat\left( k\cdot p +\omega	\right) (\bar{u}\cdot T^{a})_{\mu} \, ,\\
\raisebox{-10mm}{\vrightAdjd}&= \ii \frac y 2 c^{a} e^{\ii b\cdot k} \, \deltahat\left( k\cdot p +\omega	\right) (\bar{v}\cdot \tilde{T}^{\alpha})_{\mu} \,. 
\ea
The rule for  $\bar{\beta}(\omega)$
is obtained by reversing the arrow and replacing $(\bar u \cdot T^a)_\mu \to (T^a\cdot u)^\mu$. The same holds for the vertex propagating $\bar{\eta}(\omega)$ after exchanging $u\to v$.

To conclude this section, we fix the normalization constant ${\mathcal N}_{BA}$ appearing in \eqref{functional}.
For this  we require that, after switching off interactions, the partition function should be normalized to one. This only leaves the moduli integration
\begin{equation} 
 \oint \frac{\dd z}{2\pi \ii} \frac{e^{z \bu\cdot u}}{z^{s+1}}= \frac{(\bar{u}\cdot u)^{s}}{s!},
 \label{eq86} 
 \end{equation} 
which sets the normalization constant to
 \begin{equation}
 {\mathcal N}_{BA} = \frac{s!}{(\bar{u}\cdot u)^{s}}\frac{\tilde{s}!}{(\bar{v}\cdot v)^{\tilde{s}}} \,.
 \end{equation}

\section{Colored scalars on the worldline}
\label{colors-appendix}
Here we build up a partition function for scalar chromodynamics, following Ref.\cite{Ahmadiniaz:2015xoa}, where a worldline model has been used to evaluate 1PI diagrams in a Yang-Mills.
This leads to  an equivalent model as the one used in Ref.\cite{Shi:2021qsb}. 
We choose to keep the gauge modulus $z$ generated by the gauge fixing of the $U(1)$ symmetry in order to see how color factorization in the classical limit is realized on the worldline. As we will see below, these factors can be trivialized in  explicit calculations upon proper implementation of color factorization on the worldline. 
We consider the following partition function
\begin{equation}\label{PF}
{\cal Z} =  {\cal{N}}\int \D A \, e^{\ii S_{\text{YM}}}
\oint\frac{\dd z }{2\pi \ii}\frac{ e^{ z \bu \cdot u}       }{ z^{s+1}} 
\int 
\D \mathsf X e^{\ii S[\mathsf X;A]},
\end{equation}
where $X=(q,\beta,\bar{\beta})$. The background expanded WQFT action can be written as 
\begin{equation}
S[\mathsf X;A] = -\int_{-\infty}^{\infty}d\sigma \, \left(
\frac{1}{2}\dot{x}^{2} +\bar{\beta}_{a}\dot{\beta}^{a}+g\dot{x}^{\mu}A_{\mu}^{a} (z\bar{u}+\bar{\beta})\cdot T^{a}\cdot (u+\beta)
\right)\;, 
\end{equation}
where $s$ is the Chern-Simons coupling, needed to perform the projection on the sub-Hilbert space propagating totally symmetric tensor products of the fundamental color representation of $SU(N)$.  ${\cal N}$ is some normalization constant to be fixed later on.

Let us move to the derivation of WFRS from the interacting action.
To set up the perturbative expansion we background expand the configuration space variables (in momentum space) as usual.
We list here all of the interaction vertices needed to perform the calculations in the main text. In addition, we absorb the gauge modulus factor $z$ inside $\bar{u}$, since they always appear together in the background expansion of the color variables. The WFRs are then
\begin{align}
\raisebox{-10mm}{\vkone } &= g c^{a} e^{\ii b\cdot k}\, \deltahat(k\cdot p+\omega )\left(p^{\mu} k^{\nu} 
+\omega \eta^{\mu\nu} \right),\\
\raisebox{-10mm}{
	\begin{tikzpicture}[thick]
	\path [dashed, draw]
	(-1,-1) -- (0,-1);
	\draw[very thick] (0,-1) arc (-210:-245:2)node[right]
	{$q^\sigma(\omega_{2})$};
	\path [very thick,draw]
	(0,-1) -- (1,-1)node[right]
	{$q^\rho(\omega_{1})$};
	\path [draw,snake it]
	(0,-1) -- (0,-2)node[below]{$A^a_{\mu}(k)$};
	\draw[->,>=stealth] (-0.5,-1.2) -- (-0.5,-1.8) node[midway, left]
	{$k$};
	\coordinate (A) at (0,-1);
	\filldraw (A) circle (1.5pt);
	\end{tikzpicture}
}&=-\ii gc^a e^{\ii b\cdot k}  \hat \delta (k\cdot p +\omega_1+\omega_2) \left( \omega_1 k^\sigma \eta^{\mu \rho}+\omega_2 k^\sigma \eta^{\mu \rho}+p^\mu k^\sigma k^\rho  \right),
\end{align}
which only contains kinematical fluctuations i.e. fields related to the background expansion of the position space variable. We will also need fluctuations of the background color variables, which we  require in QCD, namely
\begin{align}
\raisebox{-10mm}{\vright } =& -\ii g e^{\ii b\cdot k} \, \deltahat\left( k\cdot p +\omega	\right) \left(  
p^\mu 
\right) (\bu\cdot T^{a})_{\sigma},\\
\raisebox{-15mm}{
	\begin{tikzpicture}[thick]
	\path [dashed, draw]
	(-1,-1) -- (0,-1);
	\draw[very thick,red] (0,-1) arc (-210:-245:2)node[right,black]
	{$\beta^\sigma(\omega_{2})$};
	\path [very thick,draw]
	(0,-1) -- (1,-1)node[right]
	{$q^\rho(\omega_{1})$};
	\path [draw,snake it]
	(0,-1) -- (0,-2)node[below]{$A^{a}_{\mu}(k)$};
	\draw[->,>=stealth] (-0.5,-1.2) -- (-0.5,-1.8) node[midway, left]
	{$k$};
	\path[](C) node[color=red,currarrow,scale=1,xscale=-1]{};
		\coordinate (A) at (0,-1);
	\filldraw (A) circle (1.5pt);
	\coordinate (B) at (0.2,-.7);
	\coordinate (C) at (.4,-.48);
\end{tikzpicture}
}=&g e^{\ii b\cdot k} \deltahat(k\cdot p+\omega_{1}+\omega_{2})\left(p^{\mu} k^{\rho} 
+\omega_1 \eta^{\mu\rho} 
\right) (\bu\cdot T^a)_{\sigma},
\end{align}
where the second  mixes kinematic and color fluctuations. The vertices propagating fluctuations of $\bar{\beta}_\sigma(\omega)$ can be obtained by reversing the arrows and replacing $(\bar u \cdot T^a)_\sigma \to (T^a\cdot u)^\sigma$ in the above ones.

\subsection*{Color factorization from the path integral}
Let us now move to color factorization in the classical limit.
First we start by fixing the normalization constant in front of the partition function \eqref{PF}.  
Absorbing divergent terms in the path integral measure and requiring the path integral to be normalized to unity when switching off the YM background allows to set the normalization constant to $\mathcal N= s! /(\bu\cdot u)^{s}$ once performed the contour integral over the unit one circle (see Eq.\eqref{eq86}),
which picks up the residue in $z=0$ of the integrand.

Let us see how  color factorization works at the level of the path integral. We consider the expectation value of two color charges on the free theory path integral. Following  \cite{delaCruz:2020bbn} we restore factors of $\hbar$ through $T^{a}\to \hbar T^{a}$. In this way, we can evaluate the above expectation value as 
\begin{align} 
\langle Q^{a}(\omega) Q^{b}(\bar{\omega}) \rangle &= {\cal N} \oint \frac{dz}{2\pi \ii} \,\frac{ e^{z \bar{u} \cdot u }}{z^{s+1} }\, \Big(
\hbar^{2}z^{2} \, (\bar{u}\cdot T^{a}\cdot u)\, (\bar{u}\cdot T^{b}\cdot u) \\
&\qquad - \frac{\ii z\hbar^{2}}{\omega}
 \bar{u}\cdot T^{a}\cdot T^{b}\cdot u - \frac{\ii z\hbar^{2}}{\bar{\omega}}
 \bar{u}\cdot T^{b}\cdot T^{a}\cdot u -\frac{\hbar^{2}}{\omega \bar{\omega}}Tr(T^{a}T^{b})
\Big)\nonumber\\
&= \hbar^{2}s(s-1) \frac{\bar{u}\cdot T^{a}\cdot u}{\bar{u} \cdot u}\, \frac{\bar{u}\cdot T^{b}\cdot u}{\bar{u} \cdot u} + \mathcal O(\hbar)\\ 
&=c^{a}\, c^{b}+ \mathcal O(\hbar) \,,
\end{align}
which lead us to define the classical color charge as
\begin{equation}
c^{a}= \hbar s \frac{\bar{u}\cdot T^{a}\cdot u }{\bar{u}\cdot u}\, .
\end{equation}
The charge  is finite in the  scaling limit in which we take $s$ to be large as $\hbar \to 0$. On the worldline side taking $s$ to be large allows us to project on large representations of the color group, when  the worldline theory is quantized.
This shows how color factorization is directly related to the power counting of the gauge modulus $z$, which we can use to keep track of the classical contributions.

\bibliographystyle{JHEP}

\renewcommand\bibname{References} 
\ifdefined\phantomsection		
\phantomsection  
\else
\fi
\addcontentsline{toc}{section}{References}
\providecommand{\href}[2]{#2}\begingroup\raggedright\endgroup

\end{document}